\documentclass[twocolumn]{aastex631}

\newcommand{\kms}{\ifmmode{\,\hbox{km\,s}^{-1}}\else {\rm\,km\,s$^{-1}$}\fi}
\shorttitle{Globular Cluster Dark Matter}
\shortauthors{Carlberg \& Keating}
\begin{document}

\title{Simulating Globular Clusters in Dark Matter Sub-Halos}

\author[0000-0002-7667-0081]{Raymond G. Carlberg}
\affiliation{Department of Astronomy \& Astrophysics \\
University of Toronto \\
Toronto, ON M5S 3H4, Canada} 
\email{raymond.carlberg@utoronto.ca}

\author[0000-0001-5211-1958]{Laura~C.~Keating}
\affiliation{Leibniz-Institut f\"ur Astrophysik Potsdam (AIP)\\
An der Sternwarte 16 \\
D-14482 Potsdam, Germany}

\begin{abstract}
A cosmological zoom-in simulation which develops into a Milky Way-like halo is started at redshift 7. The initial dark matter distribution is seeded with  dense star clusters, median mass $5\times 10^5 M_\sun$, placed in the largest sub-halos present, which have a median peak circular velocity of 25 \kms. Three simulations are initialized using the same dark matter distribution, with the star clusters  started on approximately circular orbits having initial median radii 6.8 kpc, 0.14 kpc, and, at the exact center of the sub-halos. The simulations are evolved to the current epoch at which time the  median galactic orbital radii of the three sets of clusters are 30, 5 and 16 kpc,  with the clusters losing about 2, 50 and 15\% of their mass, respectively.  Clusters started at small orbital radii have so much tidal forcing that they are often not in equilibrium. Clusters started at larger sub-halo radii have a velocity dispersion that declines smoothly to $\simeq$20\% of the central value at $\simeq$20 half mass radii.  The clusters started at the sub-halo centers can show a rise in velocity dispersion beyond 3-5 half mass radii. That is, the clusters formed without local dark matter always have stellar mass dominated kinematics at all radii, whereas about 25\% of the clusters started at sub-halo centers have remnant local dark matter.
\end{abstract}

\section{INTRODUCTION}

Globular clusters have long been known to be important tracers of galactic structure \citep{1918PASP...30...42S,1936JRASC..30..153P}. The old stellar halo and its low metal abundance globular clusters have no clear gradients in age or metallicity, which suggests that the halo was largely accreted onto the galaxy \citep{1978ApJ...225..357S}. The recognition of substantial stellar substructures and associated globular clusters in the Galactic halo bolsters the view that a collection of old, metal poor, dwarf galaxies largely built up the halo \citep{2006ApJ...642L.137B,2008A&ARv..15..145H} with globular clusters formed in the dark matter halos of pre-galactic dwarfs. 

The observational study and the physical formation mechanisms of star cluster formation is an area of vigorous research \citep{2010RSPTA.368..713L,2019ARA&A..57..227K,2020SSRv..216...69A}. The great ages of metal poor globular clusters \citep{2013ApJ...775..134V} puts observational study of their formation beyond current observational capability although that is expected to change as new telescopes come online \citep{2002ApJ...573...60C,2017MNRAS.469L..63R,2018MNRAS.479..332B}. The disk globular clusters, those having [Fe/H] $\gtrsim -1$, may well be formed as the high mass end of the observed star cluster formation processes of galactic disks \citep{1987ApJ...323...54E,2010ARA&A..48..431P}. The formation of low metallicity halo globular clusters also is likely to occur in a dark matter halo. That is, the dense, self-gravitating, gas required for star formation \citep{1968ApJ...154..891P,1984ApJ...277..470P} could arise at high redshift in an angular momentum supported disk within a dark halo \citep{1980MNRAS.193..189F} or in colliding galaxies \citep{1995AJ....109..960W}. More generally, cosmological star cluster formation simulations suggest that  any dark halo location with sufficient gas mass and pressure can lead to dense star cluster formation \citep{2016ApJ...831..204R,2018MNRAS.474.4232K,2019MNRAS.490.1714P,2020ApJ...890...18M,2020MNRAS.493.4315M}. 

Simulations and physical arguments find that the formation phase of globular star clusters is short compared to galaxy dynamical timescales \citep{2020MNRAS.494..624K} ending relatively cleanly with massive star winds and explosions expelling residual gas from the cluster. The outcome is that after a few hundred million years, globular clusters can be treated as stellar dynamical objects \citep{2018MNRAS.474.4232K}.

Globular clusters are subject to considerable dynamical evolution over their lifetime, leaving a survivor population significantly diminished from that at formation \citep{1975ApJ...196..407T,1984MNRAS.207..185C,1988ApJ...335..720A,1997ApJ...474..223G,2001ApJ...561..751F}. The characteristic densities of dark matter halos rise with redshift leading to relatively stronger tidal forces and dynamical friction. Friction also increases inversely with the dark matter halo velocity dispersion which can be important even at low redshift \citep{2001ApJ...552..572L}. If star formation primarily takes place in an angular momentum supported gas disk \citep{1980MNRAS.193..189F} tidal shredding can be a problem for globular clusters at high redshift in the inner region of sub-halos. Dark matter halo centers are a special location for the formation of globular clusters. Nuclear star clusters are at the center of galactic dark halos with many properties in common, most notably among the lower mass early type nuclear star clusters \citep{2020A&ARv..28....4N}. 

The formation sites of the old, metal poor, halo globular clusters are not currently known. The purpose of the simulations reported here is two-fold. First, to study the orbits, mass loss and mass-radius relations of globular star clusters as a function of radial distance from the center of dark matter sub-halos at high redshift.  The second purpose is to measure the stellar velocities in the outskirts of the clusters to examine how the velocity dispersion profile beyond the half-mass radius is related to the amount, if any, of residual dark matter present. These velocity profiles will be a guide for observational tests for local dark matter around globular clusters.

\section{Star Cluster Simulation Setup}

The dark matter starting point uses a FIRE simulation zoom-in, specifically model m12i that was initiated with MUSIC \citep{2011MNRAS.415.2101H}, that leads to a Milky Way-like galactic halo  \citep{2014MNRAS.445..581H,2016ApJ...827L..23W,2018MNRAS.480..800H,2019MNRAS.487.1380G}.  We evolved the FIRE particle distribution down to redshift 7 to give a starting point with a cosmological age of approximately 0.8 Gyr, closer to estimated globular cluster ages, and having lower, less computationally challenging, initial densities. The dark matter particles have a mass of $3.518\times 10^4 M_\sun$ and a softening of 40 parsec, which limits the two-body heating of the stars to 1-2 \kms. The dark matter softening radius is larger than a globular cluster which limits our ability to resolve the dark matter dynamics on the cluster scale. Nevertheless, the results from these simulations will be useful to guide other studies.

\begin{figure}
\begin{center}
\includegraphics[angle=0,scale=0.34,trim=20 20 20 70, clip=true]{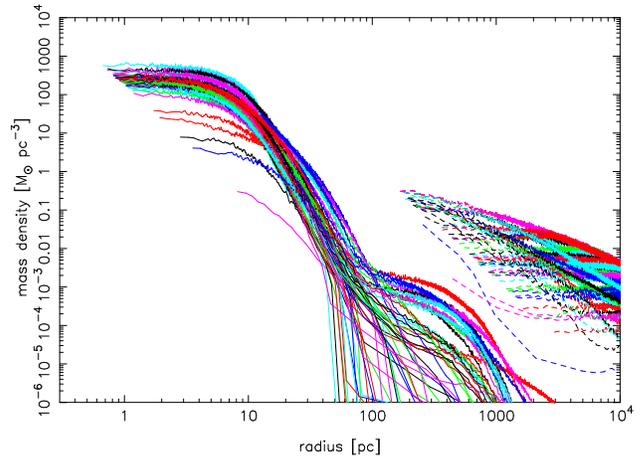}
\end{center}
\caption{
The radial mass density profiles of the star clusters (solid lines) and the dark matter (dashed lines) for the clusters placed at the centers of the dark matter halos. The measurement is made at a time of 300 Myr after the start to allow the two distributions to relax. Both stars and dark matter particles are treated as point masses for the density measurement. The line colors cycle through the clusters, matching between sub-halo and star cluster.
}
\label{fig_den516}
\end{figure}

The dark matter initial conditions provide a well understood context into which the model globular star clusters are introduced. At the start, the star particles in the clusters dominate local gravity on scales of tens of parsecs, Figure~\ref{fig_den516}. On larger scales the star clusters add so little mass that the three simulations are nearly identical to dark matter only simulations. Since the primary interest here is in the halo globular clusters the absence of a gas component, in particular a galactic disk and bulge, should not be a major factor for the dynamics of the clusters. The dark matter sub-halos in the z=7 initial distribution were located with the Amiga Halo Finder \citep{2004MNRAS.351..399G,2009ApJS..182..608K}. Rather than the conventional over-density of 200 times the critical density, an over-density of 2000 is used which gave better halo center locations for several halos, but generally found the same halos. Halos with a minimum mass of $5\times 10^8 M_\sun$ were selected as potential sites for a star cluster. 

Star clusters were generated using a King W=7 model, with cluster masses drawn from a $N(M) \propto M^{-1}$ mass function limited to the range $4-20\times 10^5 M_\sun$. The mass function is shallower than the $M^{-2}$ observed for young, massive, clusters \citep{2010ARA&A..48..431P} to allow a more uniform spread of cluster masses. The lower mass cutoff eliminates the lower mass clusters which are difficult to resolve and have sufficient internal heating that they may not survive to the final time in the strong tidal fields of the high redshift start. The total mass of star clusters assigned to a sub-halo of mass $M_{sh}$ is $\eta M_{sh}$, with $\eta = 4 \times 10^{-4}$. A halo of $1.3\times 10^{10} M_\sun$ was populated with 8 star clusters and the lowest mass sub-halos had only a single star cluster. To generate the same number of clusters at sub-halo centers required boosting the minimum sub-halo mass to $9\times 10^8 M_\sun$ and lowering $\eta$ to $1\times 10^{-4}$. The star clusters were composed of 4 $M_\sun$ star particles with a softening of 2 pc.

\begin{figure}
\begin{center}
\includegraphics[angle=0,scale=0.34,trim=20 20 20 70, clip=true]{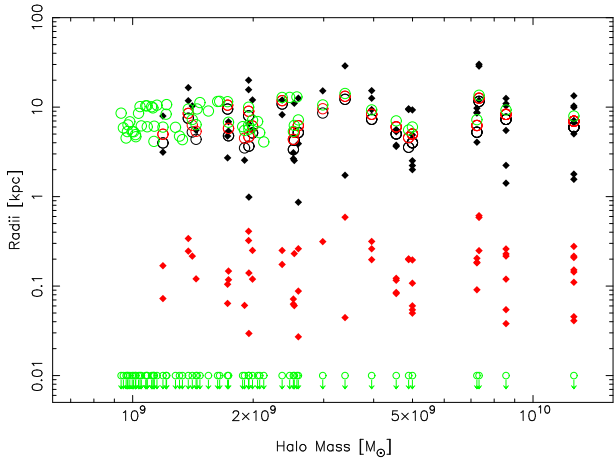}
\end{center}
\caption{
The radius of the peak of the rotation curve of the dark halos (offset vertically slightly to avoid overlap) vs their mass within over-density 2000 (open symbols) and the initial radial location of the star clusters (diamonds). The three simulations start with the same dark matter distribution. The colors are black, for the outer halo clusters; red, for the inner halo clusters; and green downward arrows, for the clusters located at halo centers. 
}
\label{fig_mrinit}
\end{figure}
%The central halo  are placed halos extending to lower masses to give nearly the same number of clusters, 62, as in the extended distributions, which have 64 clusters.

The star clusters were started on approximately circular orbits in an exponential disk distribution randomly oriented in the dark matter sub-halo. The scale radius of the disk was proportional to the radius of the peak of the circular velocity of the sub-halo, a median of 6.8 kpc for the outer clusters and  0.14 kpc for the inner clusters. The initial cluster orbital radii  and the radius of the peak of the halo circular velocity are plotted in Figure~\ref{fig_mrinit}. The clusters were assigned a tangential velocity  of the circular velocity of a Hernquist mass model  \citep{1990ApJ...356..359H} with a 2 \kms\ random velocity added in each velocity component. The star cluster outer radii were scaled to the tidal radius as determined from the local circular velocity. The star clusters placed in sub-halos used tidal radii computed as 0.002 of the radius of the peak of the halo velocity curve  and started with the center of mass velocity of the sub-halo. The adopted star particle softening of 2 pc means that the inner few parsecs of the clusters were not resolved so that most of the clusters expanded to reach equilibrium in the first 100 Myr. 

The simulations were evolved with the Gadget-4 code \citep{Gadget4}. The code is modified to incorporate gravitational collisions between stars in the clusters using the method of \citet{2018ApJ...861...69C} which adds random velocities to the stars in the shell between 1.5 to 2.0 times each cluster\rq{}s virial radius. The heating rates of massive clusters is updated on the basis of new NBODY6 \citep{1999PASP..111.1333A} star cluster simulations \citep{2021MNRAS.503.3000M}, the difference being a reduced rate of heating for clusters above $10^5 M_\sun$. The change is not very significant for these simulations since tidal heating usually dominates.

\subsection{Tides and Dynamical Friction}

\begin{figure}
\begin{center}
\includegraphics[angle=0,scale=0.34,trim=20 20 20 70, clip=true]{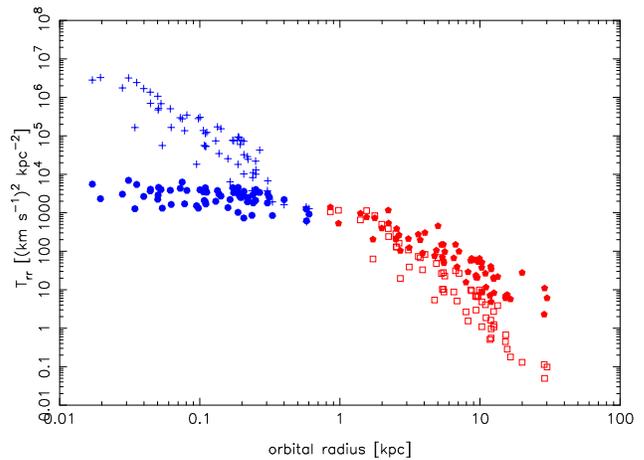}
\end{center}
\caption{
The radial tide as estimated from an approximation to the sub-halo tide (open symbols) and as measured in the simulation (solid symbols). The initial orbital radii in the small halo radius simulations are blue and the large radius  are red.
}
\label{fig_tides}
\end{figure}

The characteristic density of dark matter halos increases with redshift, $z$ approximately as $(1+z)^3$,  increasing tidal forces and dynamical friction. The Hernquist potential \citep{1990ApJ...356..359H}, $\phi = -GM/(r+a)$, is a convenient approximation to the potential inside a sub-halo of mass $M$ with its circular velocity maximum at radius $r=a$. Moreover, the distribution function for the equilibrium Hernquist sphere is used to reconstruct the dark matter distribution used for the simulations reported here. The tidal tensor is the matrix of the second derivatives of the potential,
$D^2_{ij} = \partial^2\phi/\partial x_i\partial x_j $. The trace of $D^2_{ij}$ is the matter density $4\pi G\rho(r)$.
The traceless tidal tensor is $T_{ij} = -D^2_{ij}-{1\over 3} {\rm Trace}(D^2) I$, where $I$ is the identity matrix.
The tidal force is $\sum_j T_{ij} \delta x_j$, where the $\delta x_j$ is the displacement from the center of the star cluster located within a sub-halo at $x=r$. The radial tidal force is,
\begin{equation}
F_r = 2GM \frac{(3r+a)}{3r(r+a)^3}\delta x_r,
\end{equation}
and the two tangential tidal forces are,
\begin{equation}
F_t = -GM\frac{(3r+a)}{3r(r+a)^3}\delta x_t. 
\end{equation}
For small $r$ the Hernquist potential's tidal forces diverge as $1/r$ towards the center of the sub-halo and at large $r$ the point mass tidal force is recovered. N-body realizations of halos have core radii comparable to their gravitational softening \citep{1991ApJ...378..496D} and the tidal force  goes to zero at the n-body halo centers.

The ratio of the dynamical friction time to the crossing time is roughly the ratio of the mass inside the star cluster orbit divided by the satellite mass from Equation (8.13) of \citet{2008gady.book.....B}. The outer clusters have dynamical friction times that are about 100 orbital times whereas the inner clusters have inspiral times comparable to a crossing time. At high redshift the internal substructure of the halos and the frequent merging help stave off rapid inspiral, but many of the inner clusters will be drawn into regions of high tidal fields.

\subsection{Star Cluster Internal Dynamics}

\begin{figure}
\begin{center}
\includegraphics[angle=0,scale=0.4,trim=70 100 20 60, clip=true]{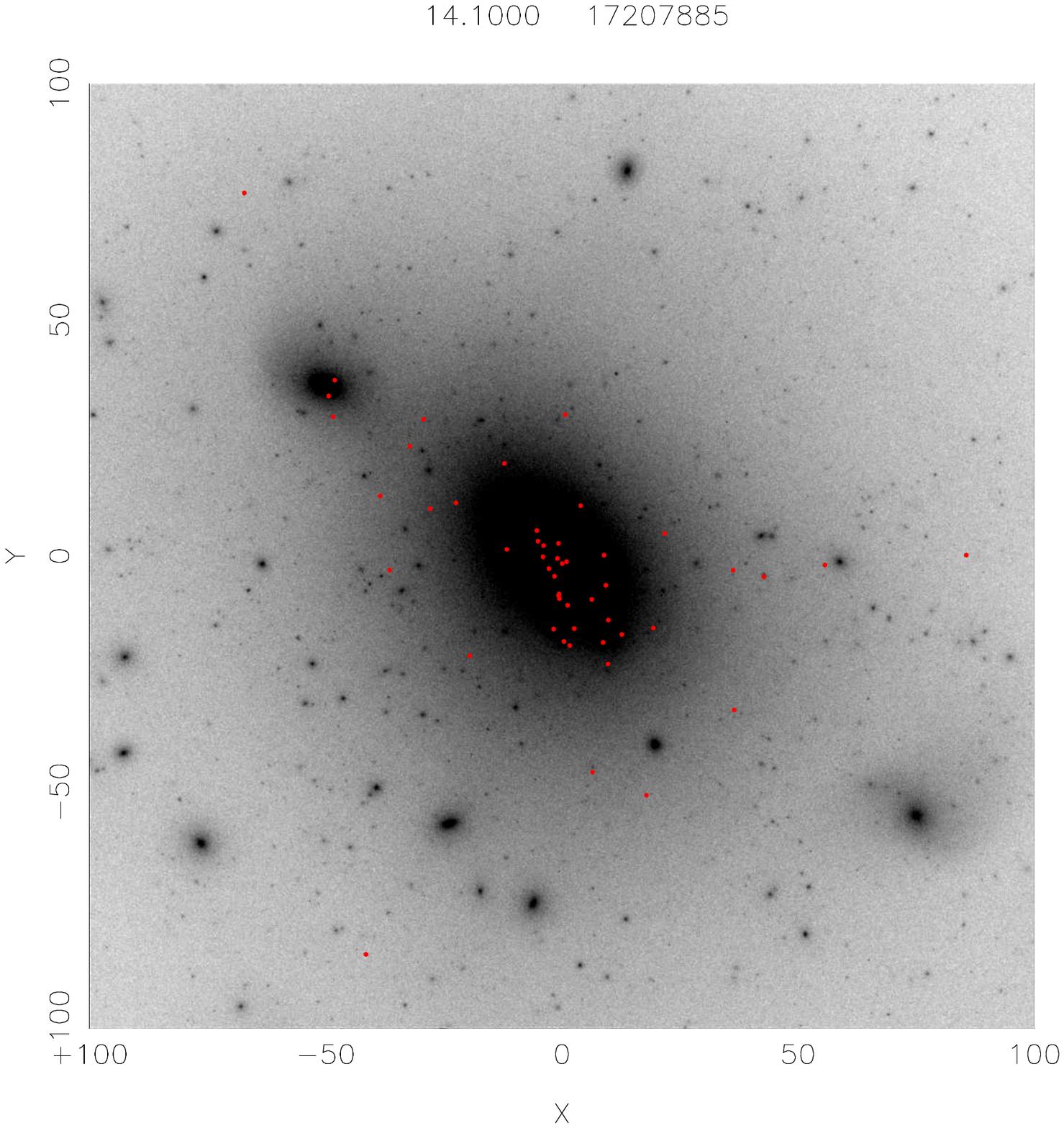}
\includegraphics[angle=0,scale=0.4,trim=70 100 20 60, clip=true]{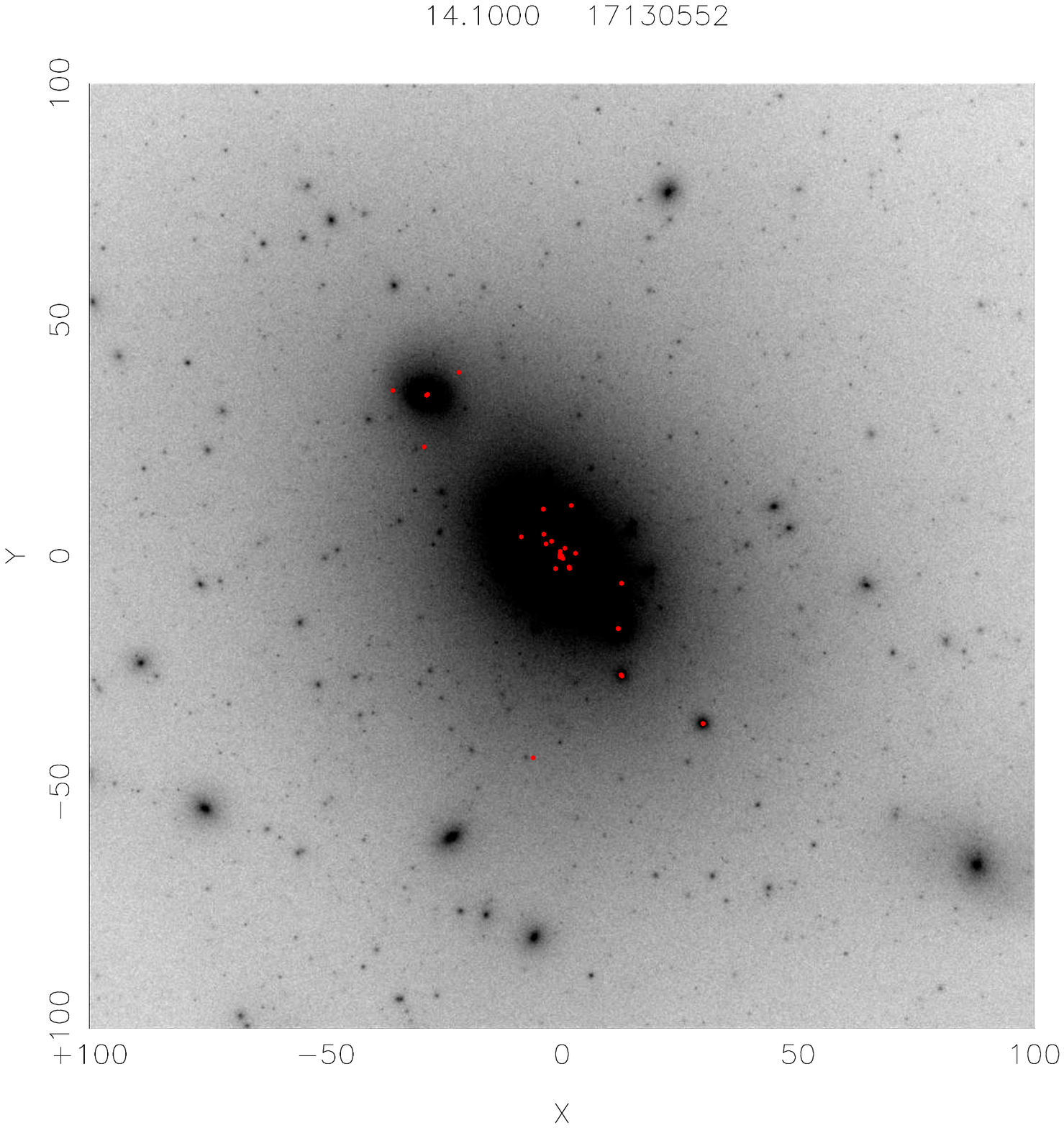}
\includegraphics[angle=0,scale=0.4,trim=70 20 20 60, clip=true]{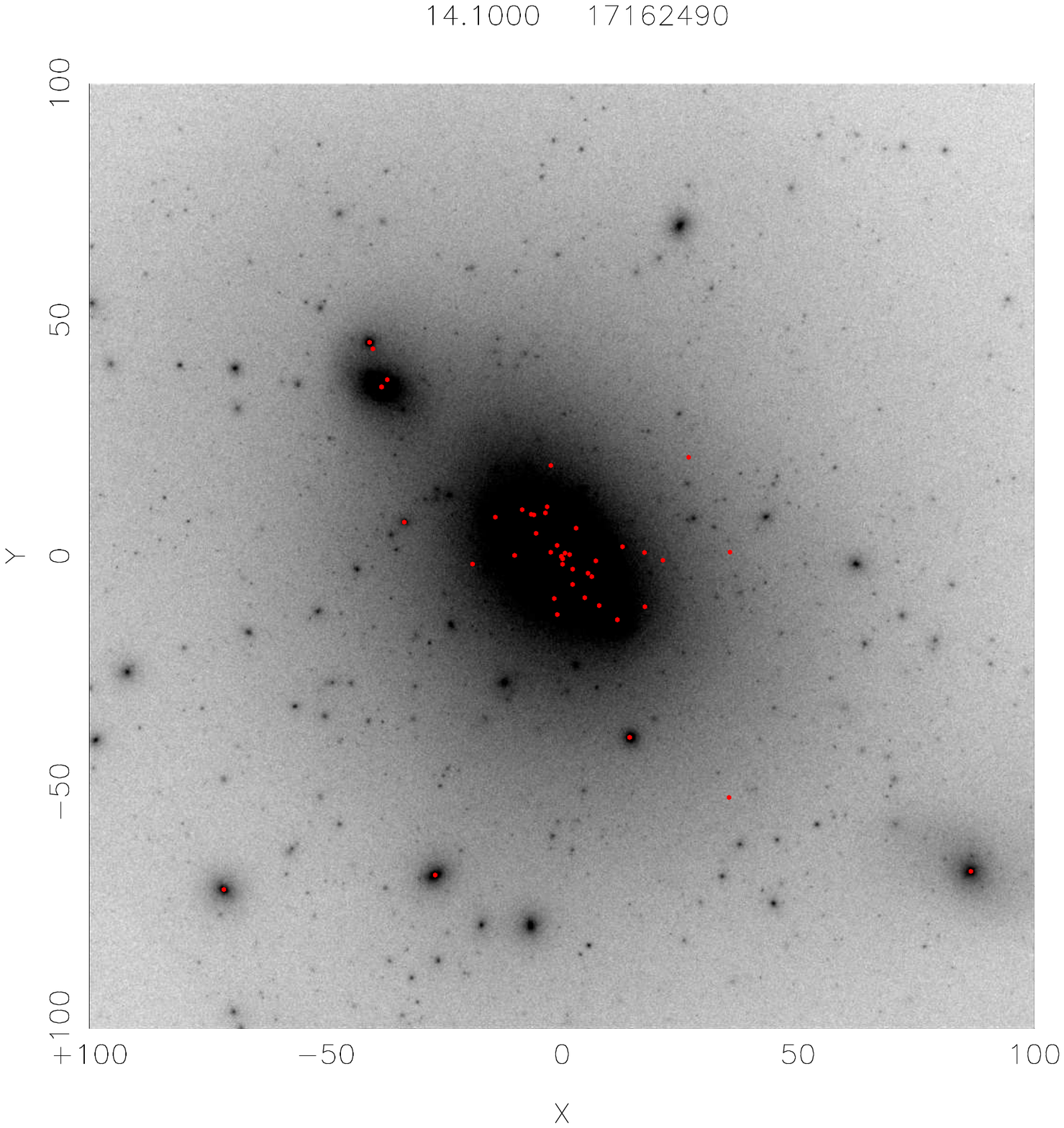}
\end{center}
\caption{The distribution of the dark matter (gray scale) and the cluster centers (red dots). The panels are from top to bottom, the large orbit, small orbit and center placement simulations. The initial small orbit and sub-halo center placements lead to a substantial fraction of the clusters spiraling to the center of the main dark matter halo. The projection is over the entire depth of the simulation.
}
\label{fig_xy}
\end{figure}

The star clusters contain about $10^5$ star particles of mass 4 $M_\sun$ having a softening of 2 pc. The relatively high mass star clusters considered used here have half mass relaxation times of $10^{9-10}$ years \citep{1987degc.book.....S} which the softening suppresses. The net effect of star-binary encounters is to extract energy from the binary giving it to field stars in the central regions of the cluster and ultimately helping to boost stars to the outer edges where tides remove them \citep{1975MNRAS.173..729H}. Tidal heating also adds energy to cluster stars, roughly in proportion to the square of the orbital radius \citep{2008gady.book.....B}. The combination of the ``kicks" and ``sweeps" \citep{2021MNRAS.503.3000M} depends on the cluster mass, half-mass radius and orbit. The clusters here do not resolve the core regions where binary interactions largely occur so a Monte Carlo scheme to add appropriate velocity increases is used \citep{2018ApJ...861...69C}. The rate of velocity change is calculated from the kinetic energy change rate evaluated at the virial radius and added to the stars that are in the shell between 1.5 and 2.0 of the  radius. The heating coefficient in the model is calibrated using NBODY6 \citep{1999PASP..111.1333A} cluster models. The heating coefficient varies slowly with star cluster mass, halving over the mass range $10^4$ to $10^{5.64} M_\sun$ above which it is kept constant. The heating routine is run every 50 Myr, intended to give the stars a chance to complete about one orbit between heating events. 

\begin{figure}
\begin{center}
\includegraphics[angle=0,scale=0.35,trim=30 30 20 80, clip=true]{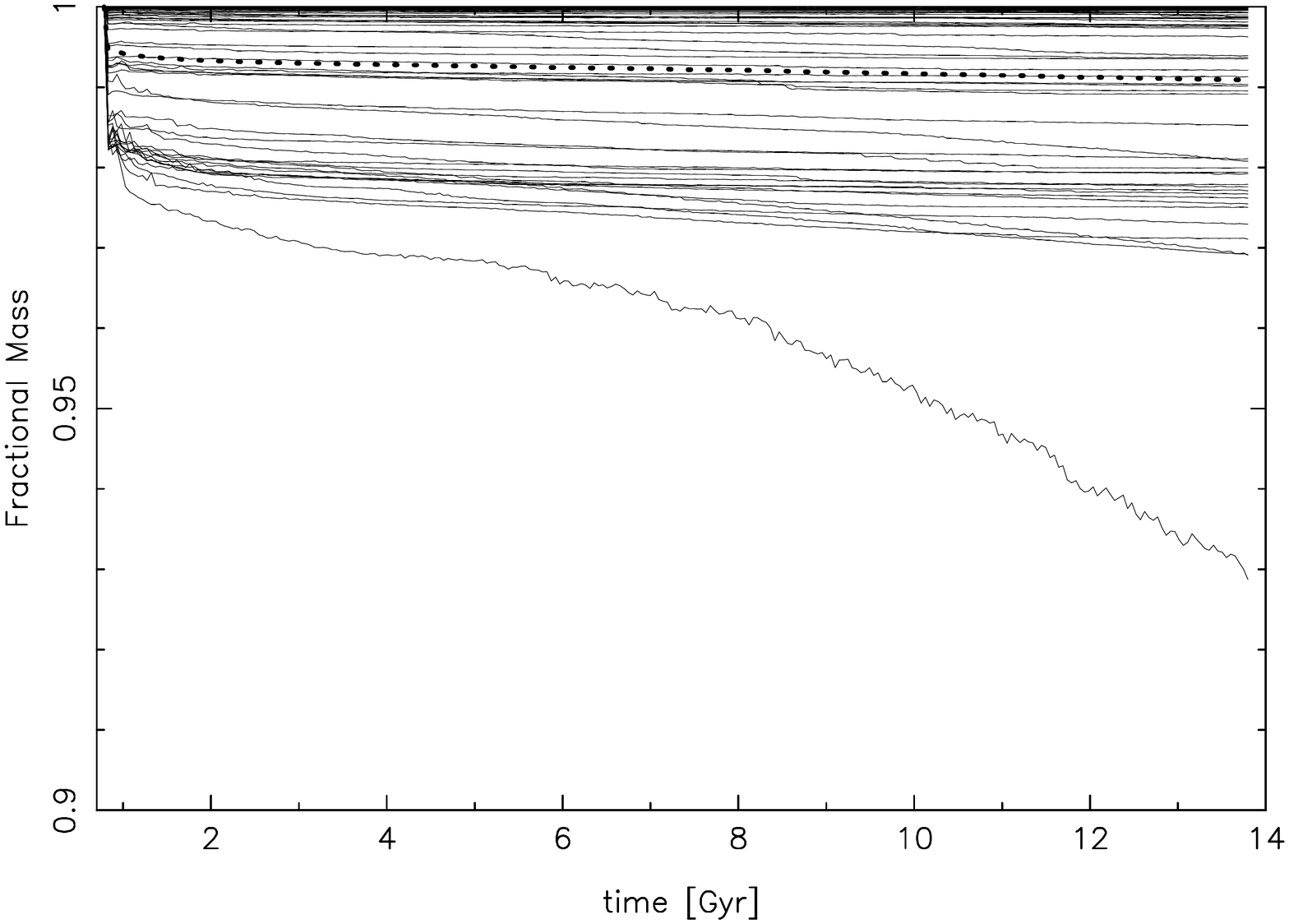}
\includegraphics[angle=0,scale=0.35,trim=30 30 20 80, clip=true]{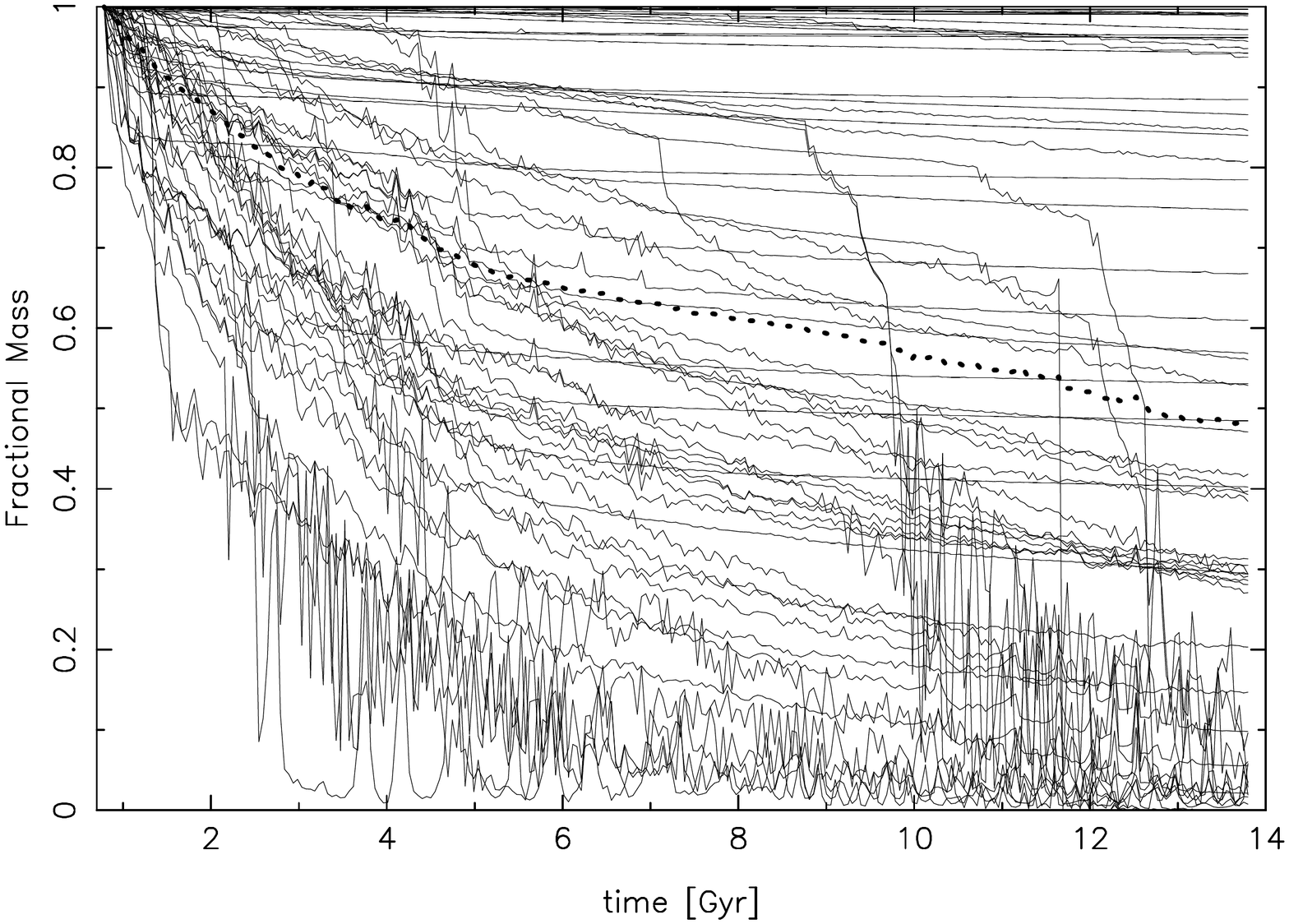}
\includegraphics[angle=0,scale=0.35,trim=30 30 20 80, clip=true]{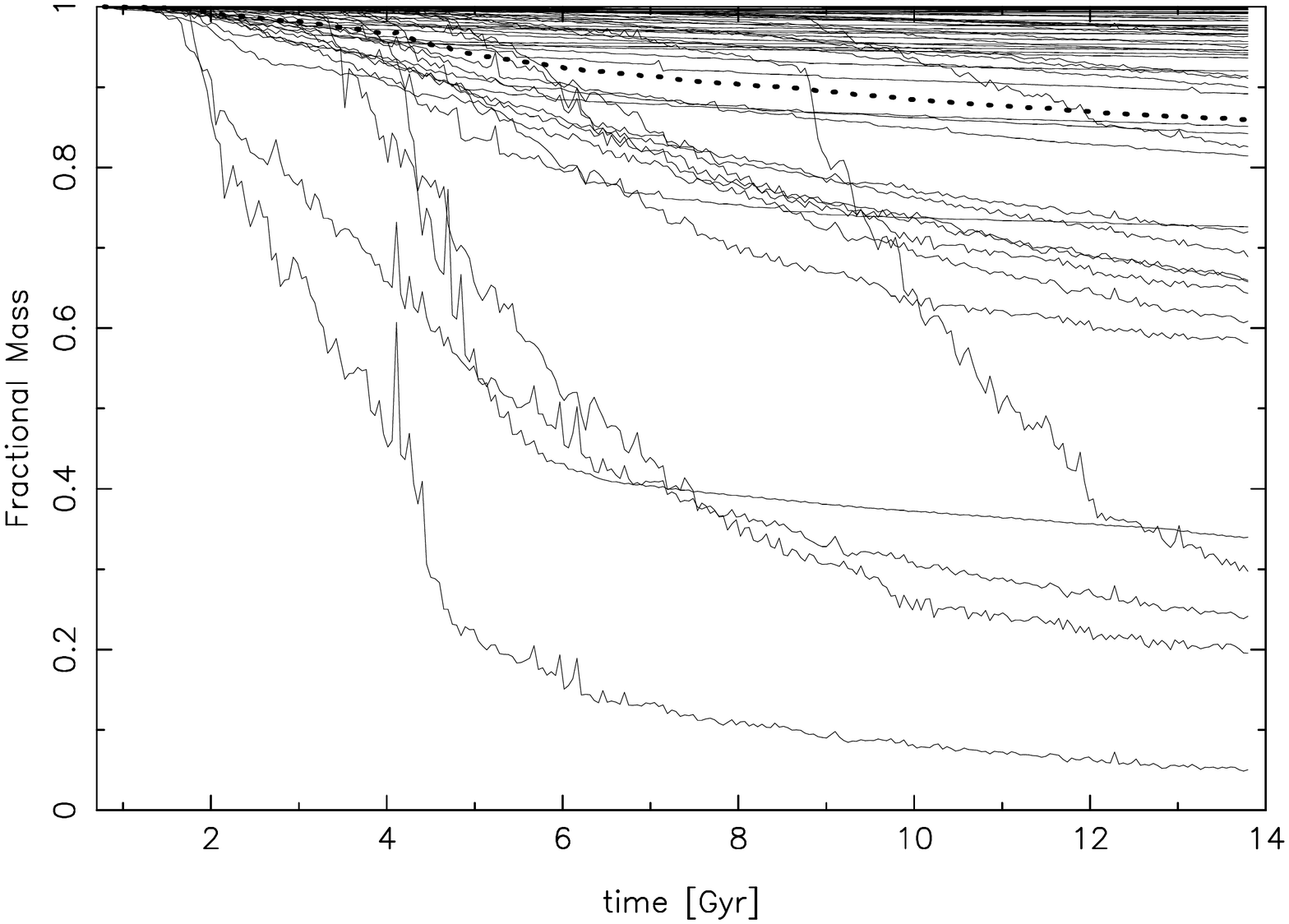}
\end{center}
\caption{
The time evolution of the clusters’ masses within a 200 pc radius. The top panel shows the clusters started at large sub-halo radii, $\approx$5 kpc,  (red points in Figure~\ref{fig_tides}), the middle panel shows the clusters started at small sub-halo radii, $\approx$0.15 kpc, (blue points in Fig.~\ref{fig_tides}) and the bottom panel shows the clusters started at halo centers. Note that the fractional mass range is 0.9-1 in the top panel, but 0-1 in the other two. The black dotted line shows the mean fractional mass with time.
}
\label{fig_mt}
\end{figure}

\subsection{Dynamical Evolution of the Simulation}

The  simulations, each composed of approximately $7\times 10^7$ dark matter particles and $10^7$ star particles, are evolved using Gadget4 \citep{Gadget4} augmented with the star cluster heating routine. The dark matter particles have a softening of 40 pc and the star particles 2 pc softening. Convergence tests in a fixed potential found that the standard value of the Gadget step size parameter {\it ErrTolInt}, 0.01-0.02 had to be reduced to a value of 0.0025 to ensure that the clusters did not expand due to numerical integration errors in the cosmological environment which has rapidly fluctuating tidal fields. With {\it ErrTolInt} of 0.0025 clusters with a half mass radius larger than approximately 8 pc $M_c/(10^6 M_\sun)$ have the same results with {\it ErrTolInt} reduced to 0.001. The presence of the star clusters completely dominates the speed of the code. The time steps settle down to about $10^{-5}$ Gyr, meaning that there are about $10^6$ time steps in a run. The simulations are usually run using 16 hosts each with 40 cores. Snapshot files are output every 10 Myr. The simulation particle masses and softening used here are just barely adequate to resolve the star clusters. In future work the particle masses should be reduced along with a reduction in the softening lengths. 

\begin{figure}
\begin{center}
\includegraphics[angle=0,scale=0.35,trim=30 10 20 80, clip=true]{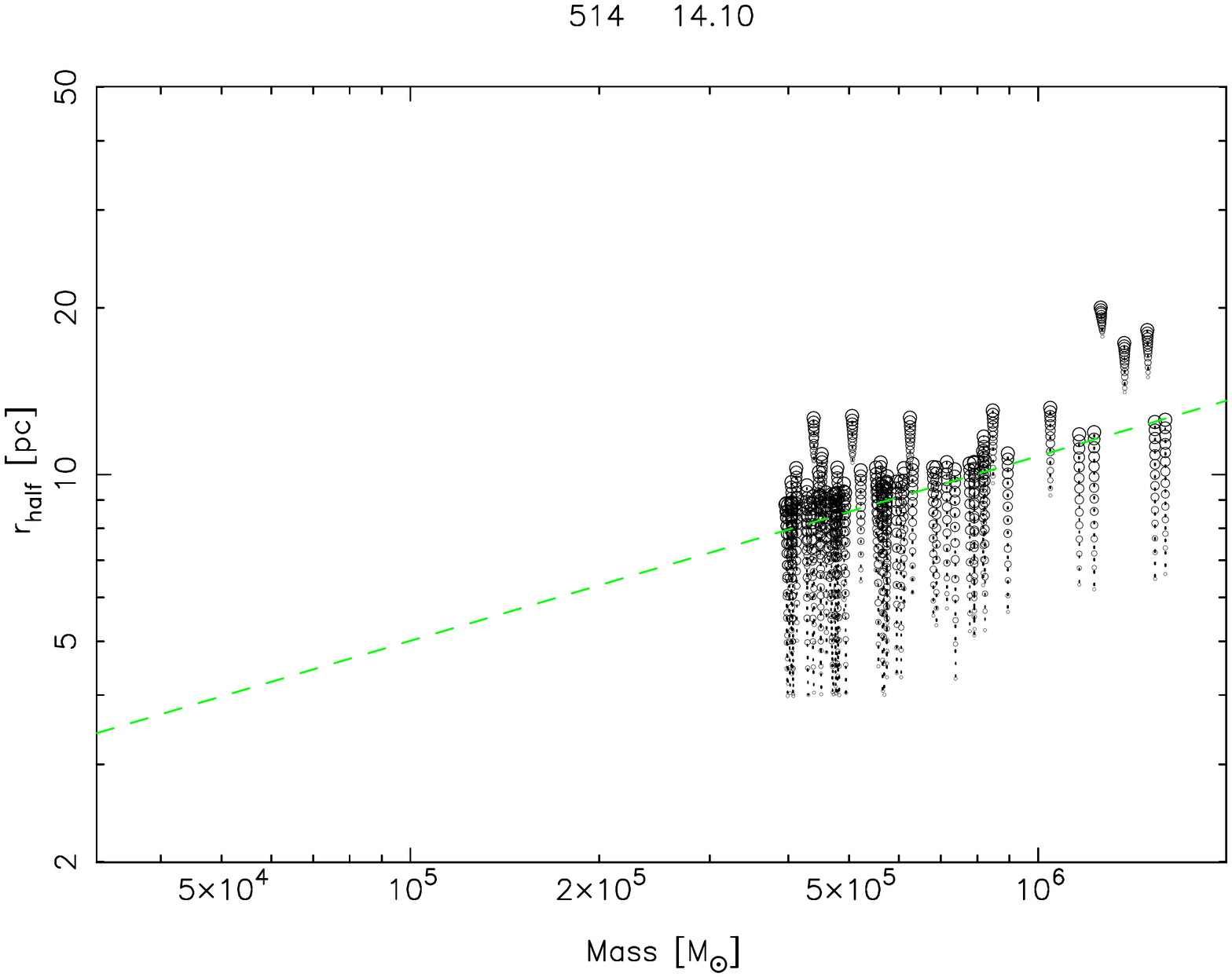}
\includegraphics[angle=0,scale=0.35,trim=30 10 20 80, clip=true]{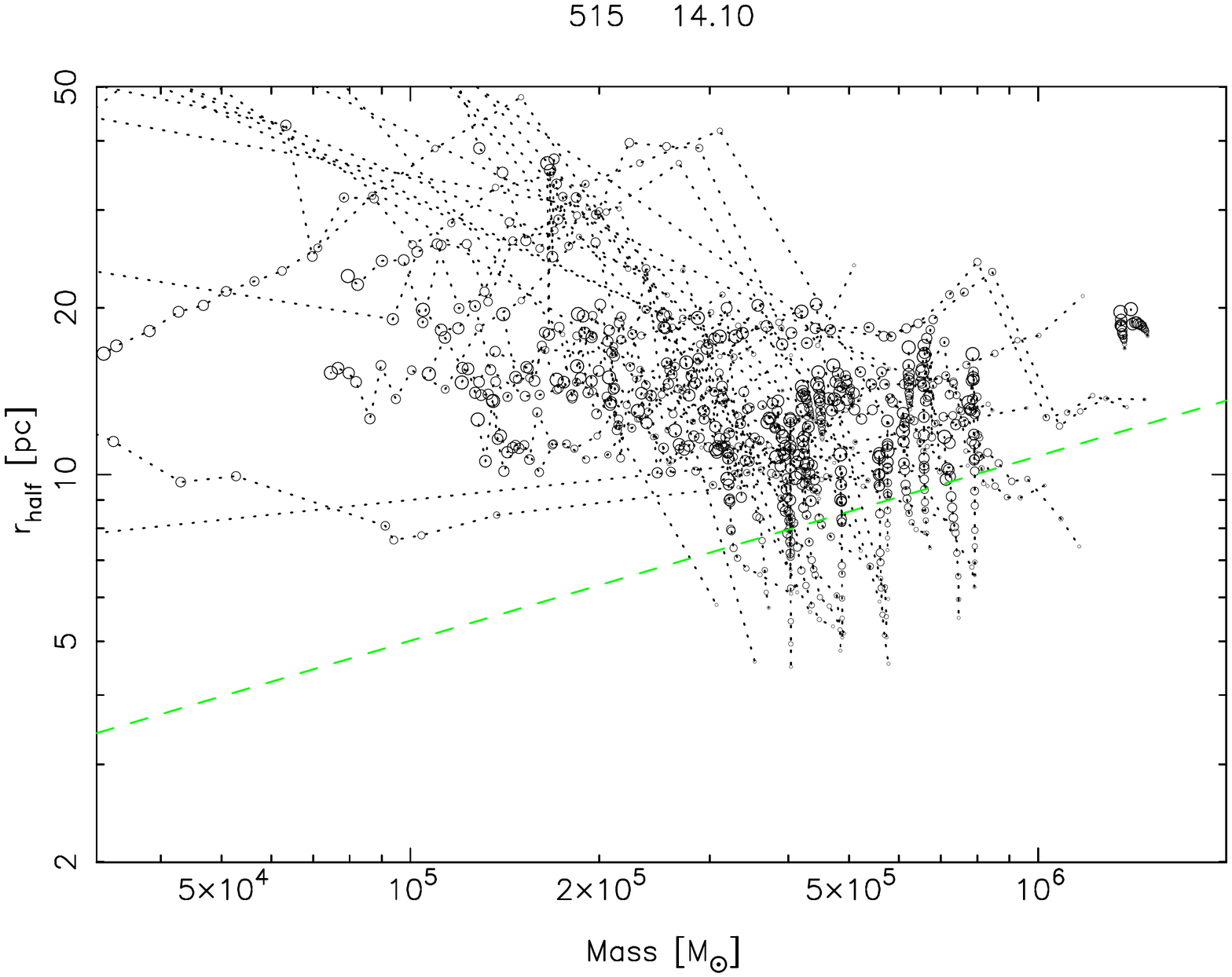}
\includegraphics[angle=0,scale=0.35,trim=30 10 20 80, clip=true]{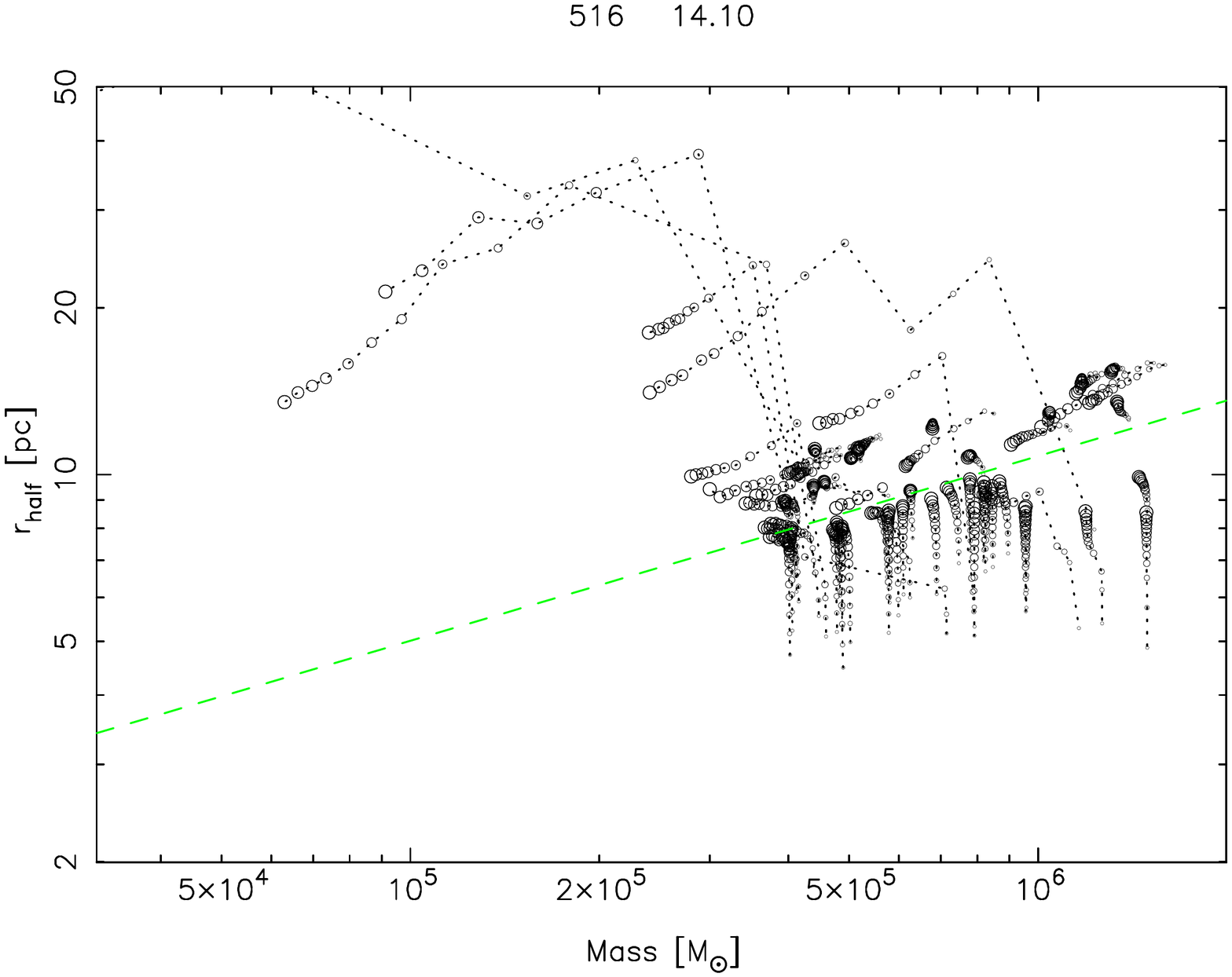}
\end{center}
\caption{The evolution of the cluster's half-mass radii every Gyr, with plotted circle size increasing with time. The green dashed line is $r_{\rm half}= 5~{\rm pc} (M_c/10^5 M_\sun)^{1/3}$. The top panel is for the large orbit cluster simulation, the middle for the small orbit simulation and the bottom is for the clusters initiated at the centers of sub-halos.
}
\label{fig_mrt}
\end{figure}

\section{Simulation Results}

The spatial evolution of the dark matter distribution and the star cluster centers is shown for the 3 simulations in Figure~\ref{fig_xy}. The red dots are the locations of the cluster centers. The median distance from the halo center is 30, 5 and 15 kpc for the simulations with clusters started around 5 kpc, 0.1 kpc, and the sub-halo centers, respectively, for clusters within 150 kpc of the center of the main halo. 

The decline of the masses of the clusters with time is shown in Figure~\ref{fig_mt} for the three simulations, with the clusters started around 5 kpc at the top, 0.15 kpc, in the middle, and the sub-halo centers at the bottom. Note that the scale for fractional mass only goes to 0.9 in the top panel, whereas the scales go to zero in the other two. The clustered started at small radii, $\approx$0.15 kpc, (middle panel) lead to mass loss of about 50\%, 15\% mass loss for the sub-halo center starts (bottom panel), and about 1-2\% for the clusters started at $\approx$5 kpc (top panel).  The survival probability is less than 50\% for massive dense star clusters started in  $\approx$0.15 kpc of the center, but $\gtrsim$90\% of those started at either at the sub-halo centers or at $\approx$5 kpc survive with less than 10\% of their mass lost.

The star clusters have clear patterns of mass-size evolution, as shown in Figure~\ref{fig_mrt}. The clusters begin with a half mass radius sufficiently small that they are strongly self-gravitating. All clusters have a slow expansion of the half mass radius with time as a result of internal and tidal heating. In the case of the clusters started around $\approx$5 kpc the tidal fields are comparatively weak and  do not remove much mass. However, the clusters started around $\approx$0.15 kpc expand and lose mass rapidly. Once they expand to a half mass radius in the 10-20 pc range, every pericenter passage heats stars which then drift to the tidal radius and pulled away at the next pericenter. The most extreme case of mass loss is a cluster that loses about 50\% of its mass in a single orbit.  There is one cluster in the $\approx$5 kpc starts (top panel) that shows a much more rapid mass loss than all the others. The random start routine placed its initial radial orbit location about 0.08 kpc from the center meaning its evolution is comparable to the set of clusters started within $\approx$0.15 kpc of their sub-halo centers shown in the bottom panel.

The star clusters started at the exact centers of the sub-halo initially have a tidal compression from the dark matter gravitational field. However, as the sub-halos merge together to build up the galactic halo,  the cluster centers orbit off center and are subject to disruptive tides until they can sink back to a new center. In the simulation here 4 of the 62 clusters gets sufficiently far from the sub-halo center that they are in the strong tide region, being at the same radii and losing mass in the same way as the clusters in the simulation started at small, $\approx$0.15 kpc, sub-halo radii. A single simulation provides only small number statistics, but the outcomes should be generally true. 

The star clusters are adequately resolved for our main purposes, although a smaller star particle softening and more stars would be very welcome. Figure~\ref{fig_den516} shows that the central star clusters are self-gravitating even if the dark matter particles were increased in number with smaller softening. The dark matter particles have softening lengths about a decade large than a star cluster half mass radius which leads to an underestimate of the dark matter gravitational field for clusters in and near the center, as is visible in the difference between the measured and analytic tidal forces shown in Figure~\ref{fig_tides}. 

\begin{figure}
\begin{center}
\includegraphics[angle=-90,scale=0.36,trim=70 120 20 60, clip=true]{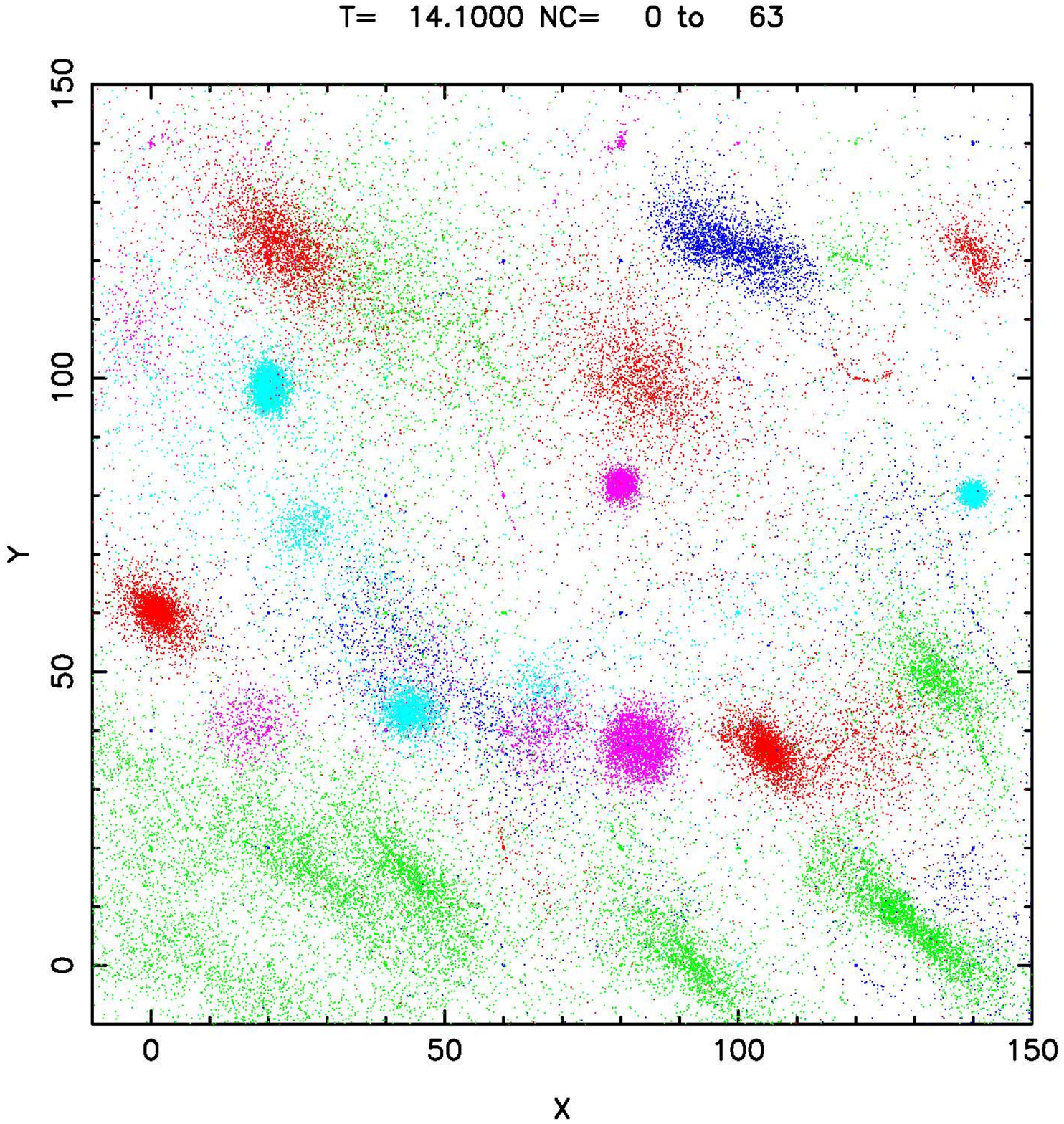}

\includegraphics[angle=-90,scale=0.36,trim=70 120 20 60, clip=true]{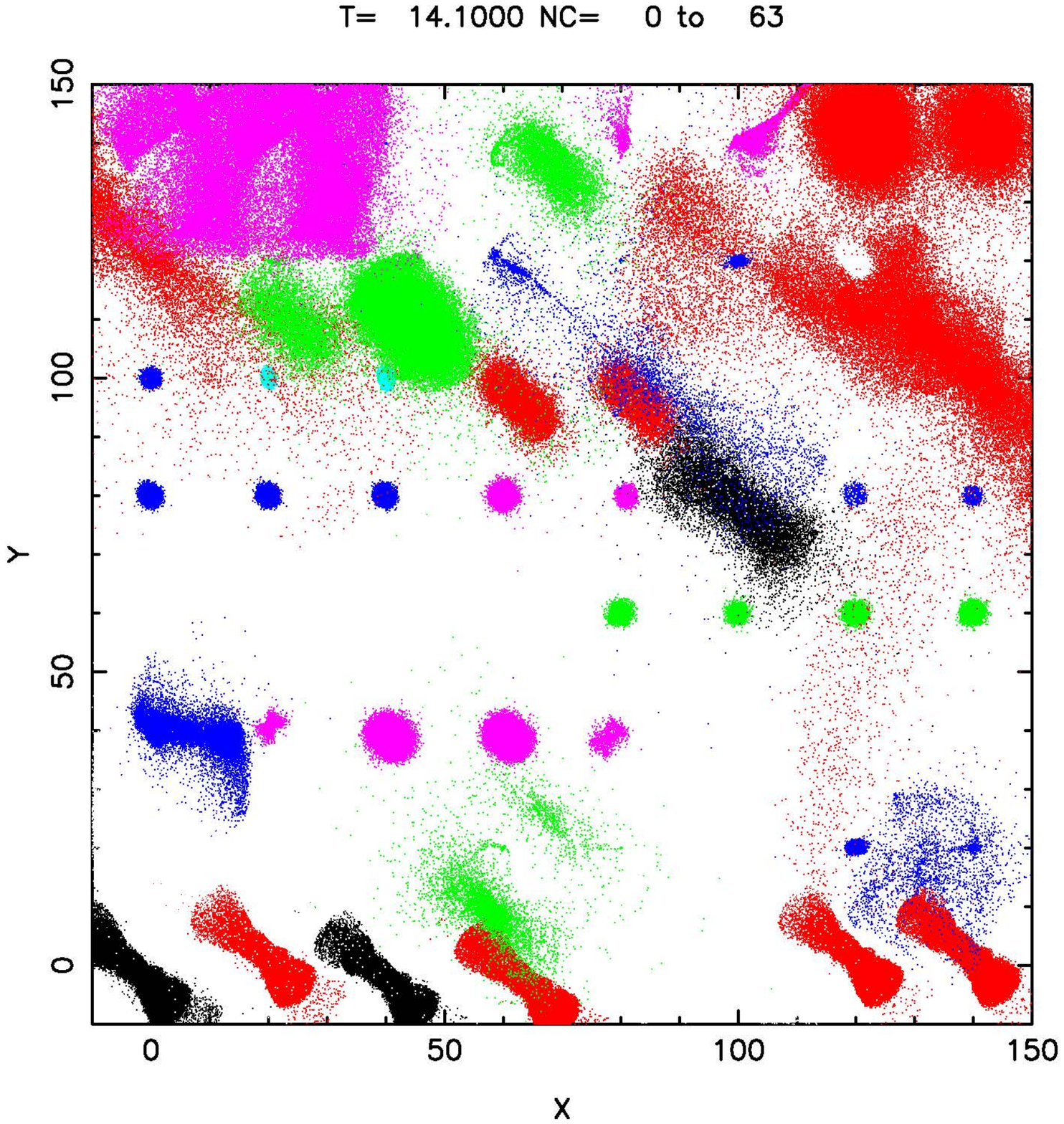}

\includegraphics[angle=-90,scale=0.36,trim=70 120 20 60, clip=true]{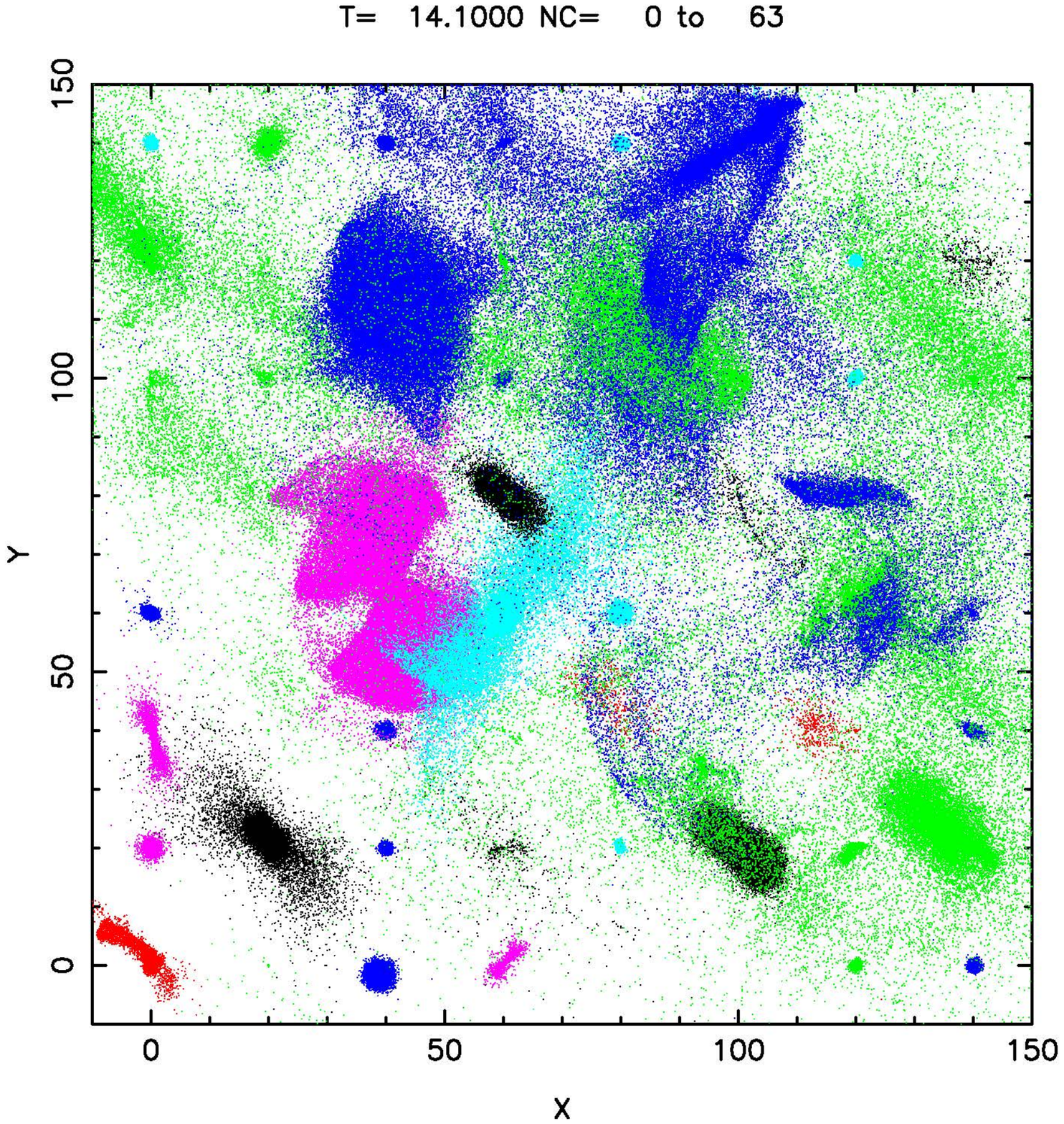}
\end{center}
\caption{The distribution of the cluster stars plotted with each cluster's center placed on a grid. The panels are from top to bottom, the large orbit, small orbit and center placement simulations. The colors indicate the final galactic radii of the cluster center: black within 3 kpc, red within 10 kpc, green within 31.6 kpc, blue within 100 kpc, cyan within 300 kpc and magenta for clusters in external sub-halos. If no cluster is visible, it has been completely dispersed, as is most noticeable in the middle panel.
}
\label{fig_cxy}
\end{figure}

Figure~\ref{fig_cxy} shows the current x-y positions of  the stars of each cluster, plotted relative to the center of mass of the cluster. Each star is plotted so it is not possible to compare densities in crowded regions. The colors indicate the distance from the center of the main halo at the final time. The central tilted bar shape reflects the recent merger history of the halo. The clusters within 3 kpc of the center have lost most of their stars which are now completely blurred out over their orbits. Most clusters more than a few kpc from the center of the main halo center have  a thin stellar stream of recently removed stars. 

\section{Star Cluster Kinematics}

\begin{figure}
\begin{center}
\includegraphics[angle=0,scale=0.35,trim=30 30 10 80, clip=true]{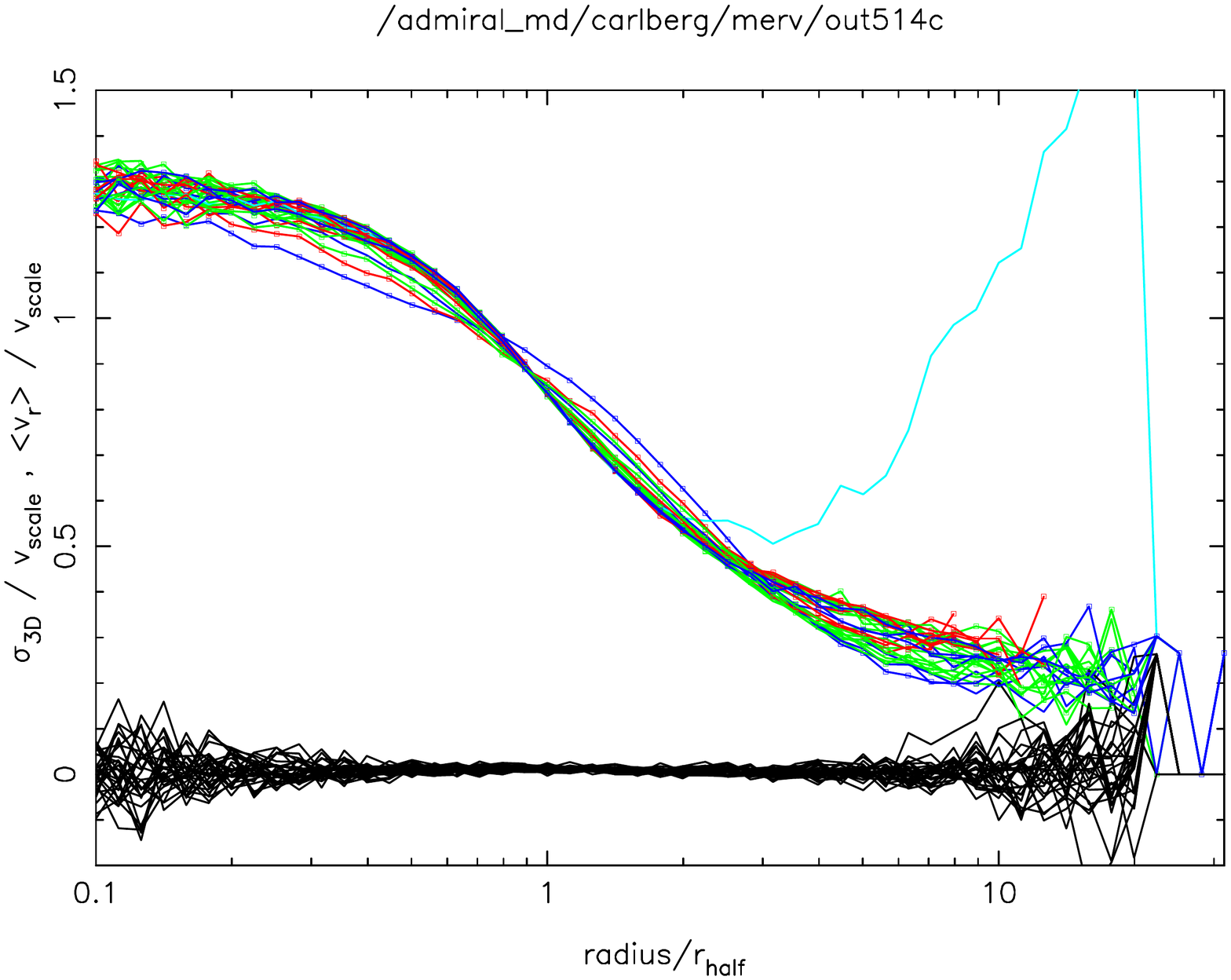}
\includegraphics[angle=0,scale=0.35,trim=30 30 10 80, clip=true]{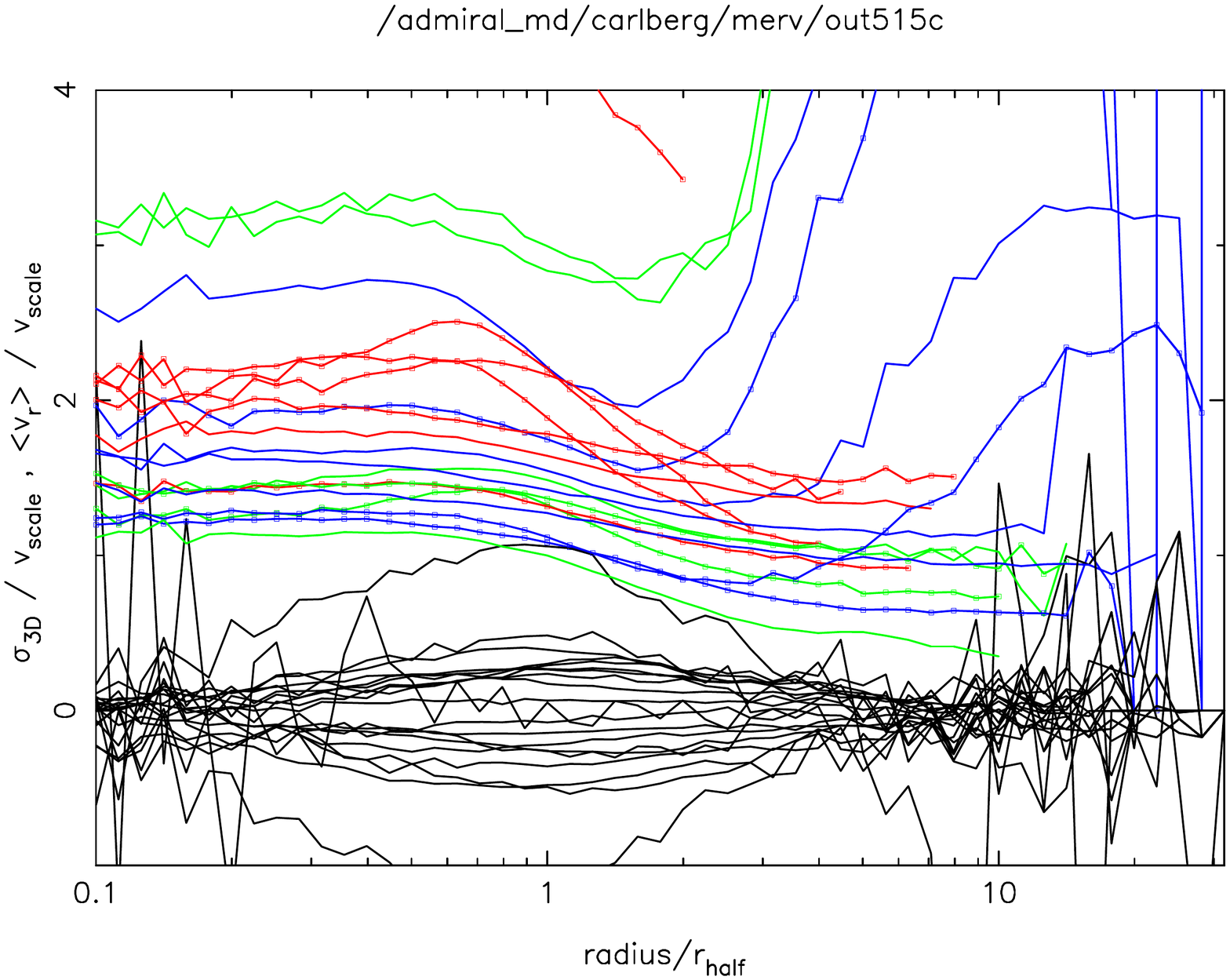}
\includegraphics[angle=0,scale=0.35,trim=30 30 10 80, clip=true]{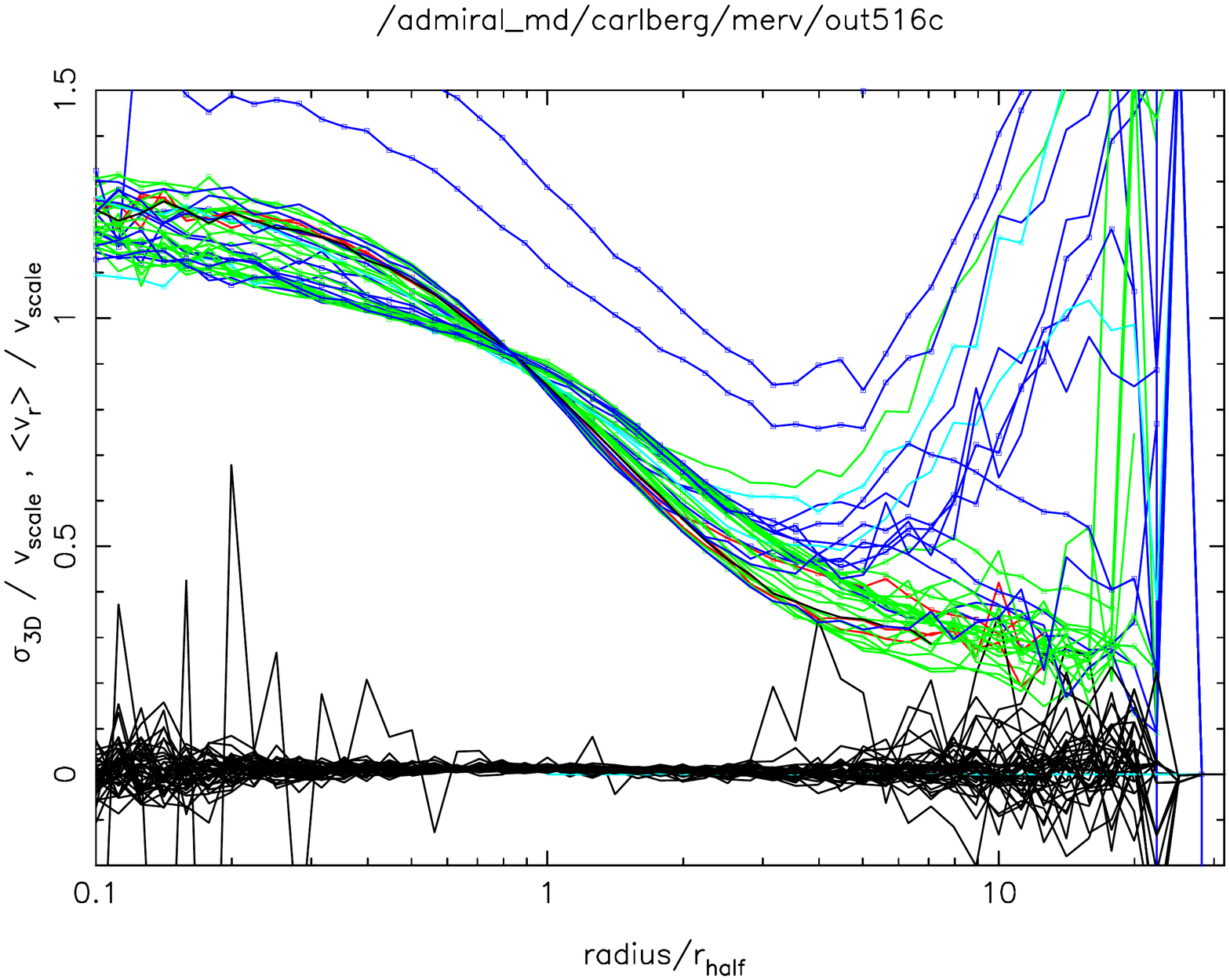}
\end{center}
\caption{The total velocity dispersion as a function of radius (within the tidal radius) for the clusters that are in the range of 3 to 150 kpc of the galactic center. The panels are in the same order as in Fig.~\ref{fig_cxy}.  The radii are normalized to each cluster\rq{s} half mass radius. The velocities are normalized to the circular velocity at the half mass radius.  The colors are the same as in Figure~\ref{fig_cxy}. The black lines near the bottom of the plot show the mean radial velocity.  The clusters moving outwards on their galactic orbit have boxes on the lines. 
}
\label{fig_vc}
\end{figure}

The simulations have two sets of long-lived clusters with limited mass loss: clusters formed either in the range of $\approx$5 kpc of the initial sub-halos, or, formed at the exact center of sub-halos, with end-point median galactic orbital radii of 30 and 15 kpc, respectively. These star clusters  have sizes comparable to the massive globular star clusters in the Milky Way. The simulated clusters have a range of orbits from the distant reaches of the outer halo to near the galactic center. Star clusters with the same mass and size appear at any galactic radius which is a key property of halo clusters \citep{1978ApJ...225..357S}.  The star clusters formed deep within their initial sub-halos, $\approx$0.15kpc, expand to half-mass radii of $\simeq$20 pc and then lose mass rapidly. 

\begin{figure}
\begin{center}
\includegraphics[angle=0,scale=0.35,trim=30 30 10 80, clip=true]{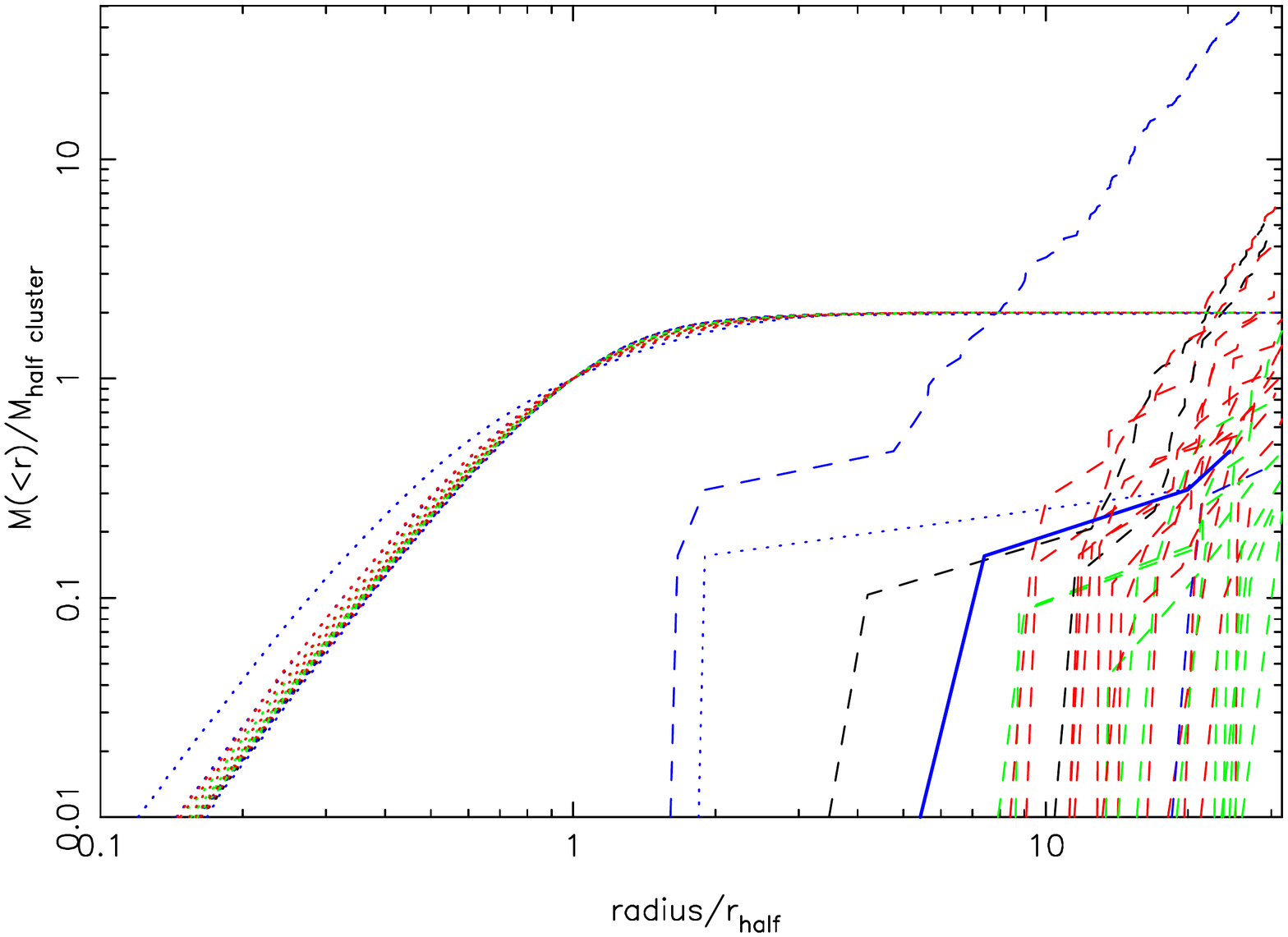}
\includegraphics[angle=0,scale=0.35,trim=30 30 10 80, clip=true]{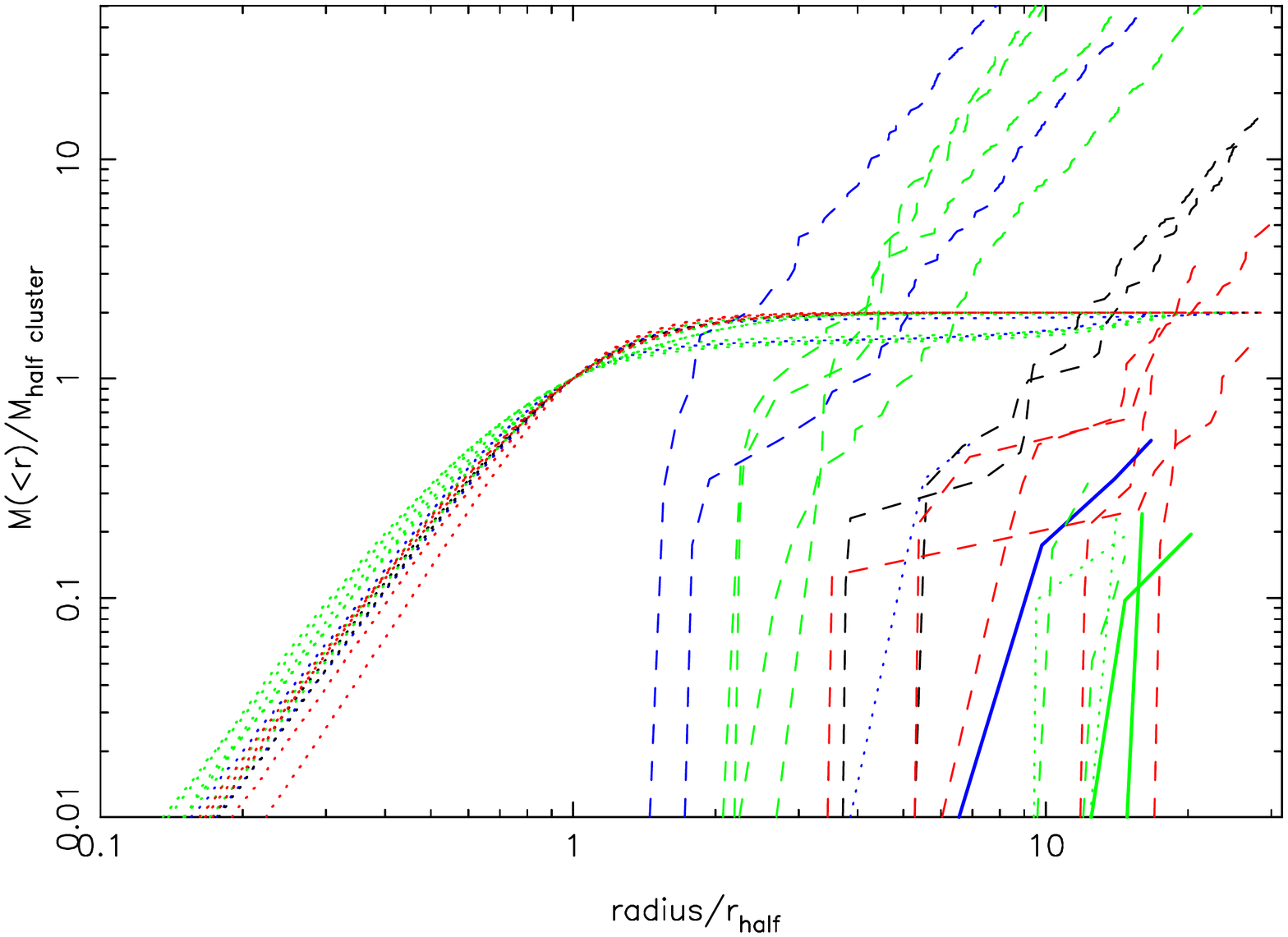}
\includegraphics[angle=0,scale=0.35,trim=30 30 10 80, clip=true]{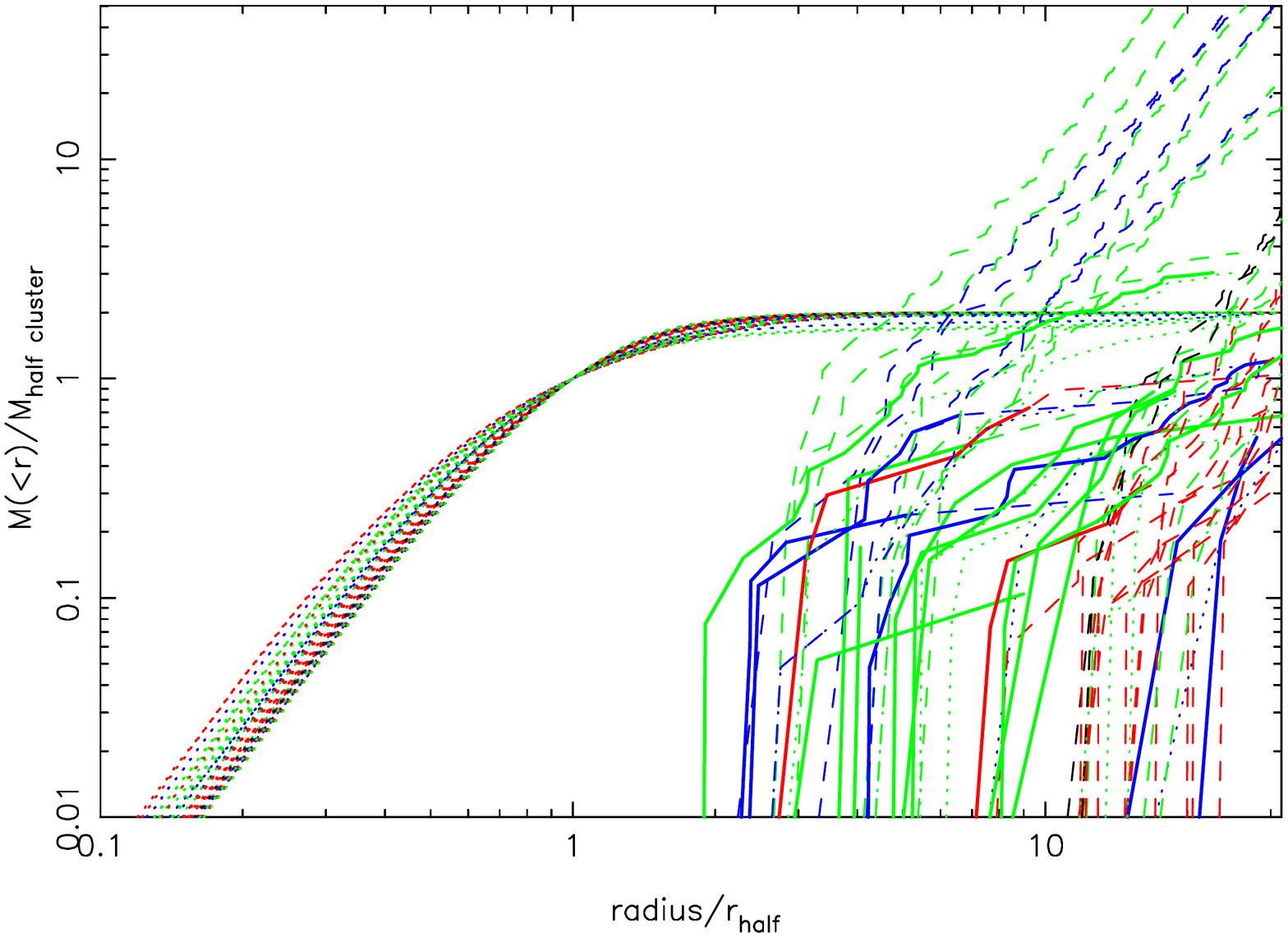}
\end{center}
\caption{The integrated particle mass as a function of radius for the same clusters and colors as in Fig.~\ref{fig_vc}, those located between 3 and 150 kpc. The particles are treated as point masses. The stellar and dark matter mass profiles are shown as dotted and solid lines, respectively. The colors are as in Fig.~\ref{fig_vc}.  Solid lines are for dark matter bound to the cluster, dashed lines are all dark matter.
}
\label{fig_msd}
\end{figure}

The kinematics of stars in the cluster outskirts, is where differences appear between the clusters orbiting in the background dark matter and the clusters that are embedded at the center of remnant sub-halo that is itself orbiting within the background dark matter of a galactic halo. Figure~\ref{fig_vc} shows the total velocity dispersion vs radius for the three simulations. The velocity curves end at the tidal radius, $r_t$, of the cluster of mass $M_c$ at galacto-centric radius $R_{GC}$ as $r_t=R_{GC}[GM_c/(2V_c^2R_{GC})]^{1/3}$. The main halo has a mass of $1.03\times 10^{12} M_\sun$ inside 251 kpc, with $V_c$ within about 10\% of 170 \kms\ over the range 15 to 150 kpc. The clusters will have unbound stars within the tidal radius, so stars with velocities greater than the central escape velocity to the scale velocity, a factor of 2.28, based on the values for a Plummer sphere, are not included in the analysis. 

The star clusters started  at $\approx$5 kpc have a velocity dispersion that drop steadily to about 15-25\% of the central value at 10-30 times the half mass radius.  There is one cluster that shows a steep rise in velocity dispersion beyond about 3 half mass radii. The random radius generated for this this cluster placed at only 0.08 kpc from the center of its own sub-halo, the smallest in the simulation. At the final epoch the cluster is at 106 kpc from the main halo center in a sub-halo. Although this is a somewhat rare occurrence, it is interesting to note. 

\begin{figure*}
\begin{center}
\includegraphics[angle=0,scale=0.13,trim=0 1000 0 0, clip=true]{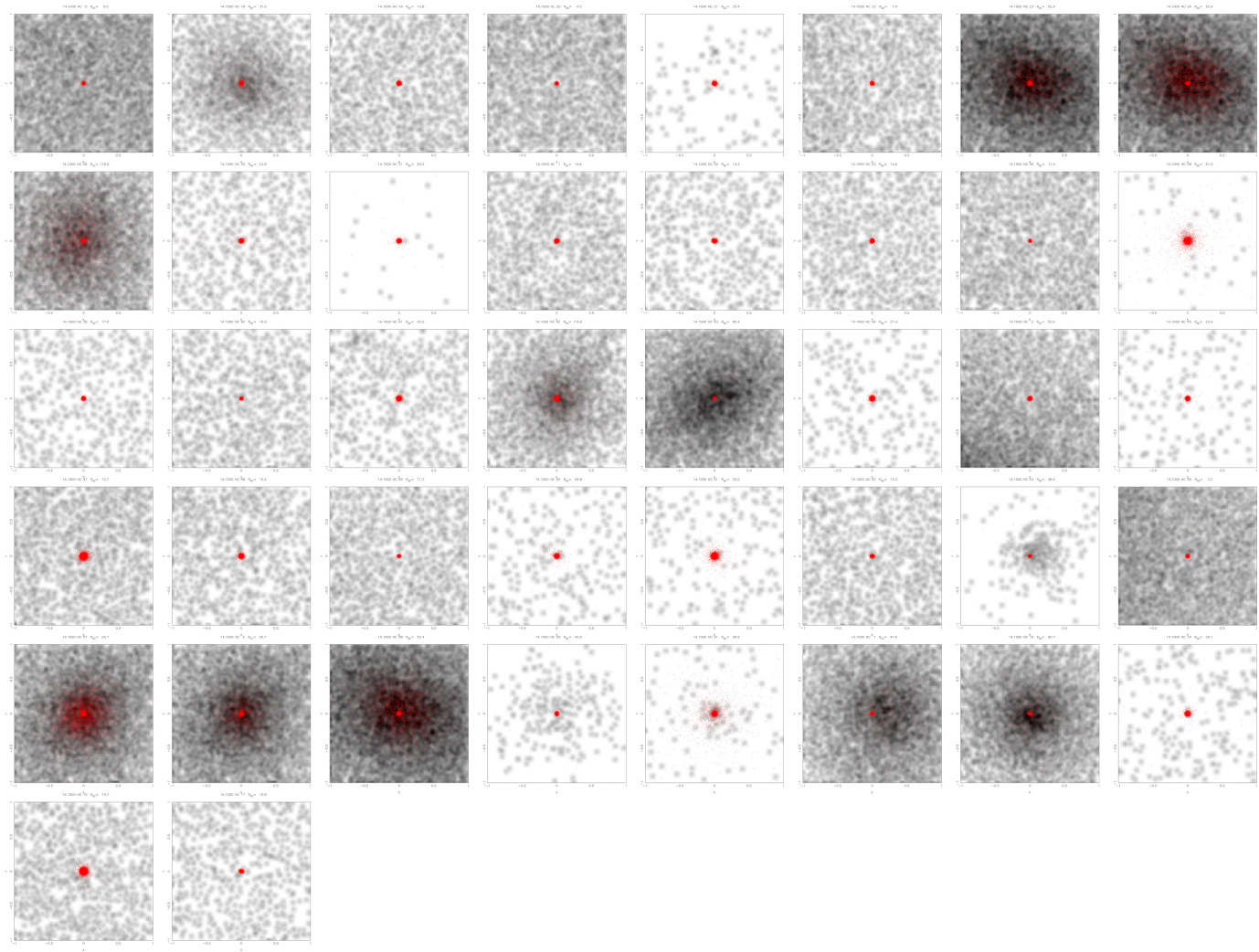}
\end{center}
\caption{Postage stamps, $\pm$1 kpc of each cluster center, for the clusters that were started at dark matter sub-halo centers. Clusters are selected to have final epoch galactic radii between 3 and 150 kpc. The dark matter density is shown as a grey scale, the red dots are every tenth star particle.
}
\label{fig_montage}
\end{figure*}

\begin{figure}
\begin{center}
\includegraphics[angle=0,scale=0.35,trim=30 30 10 70, clip=true]{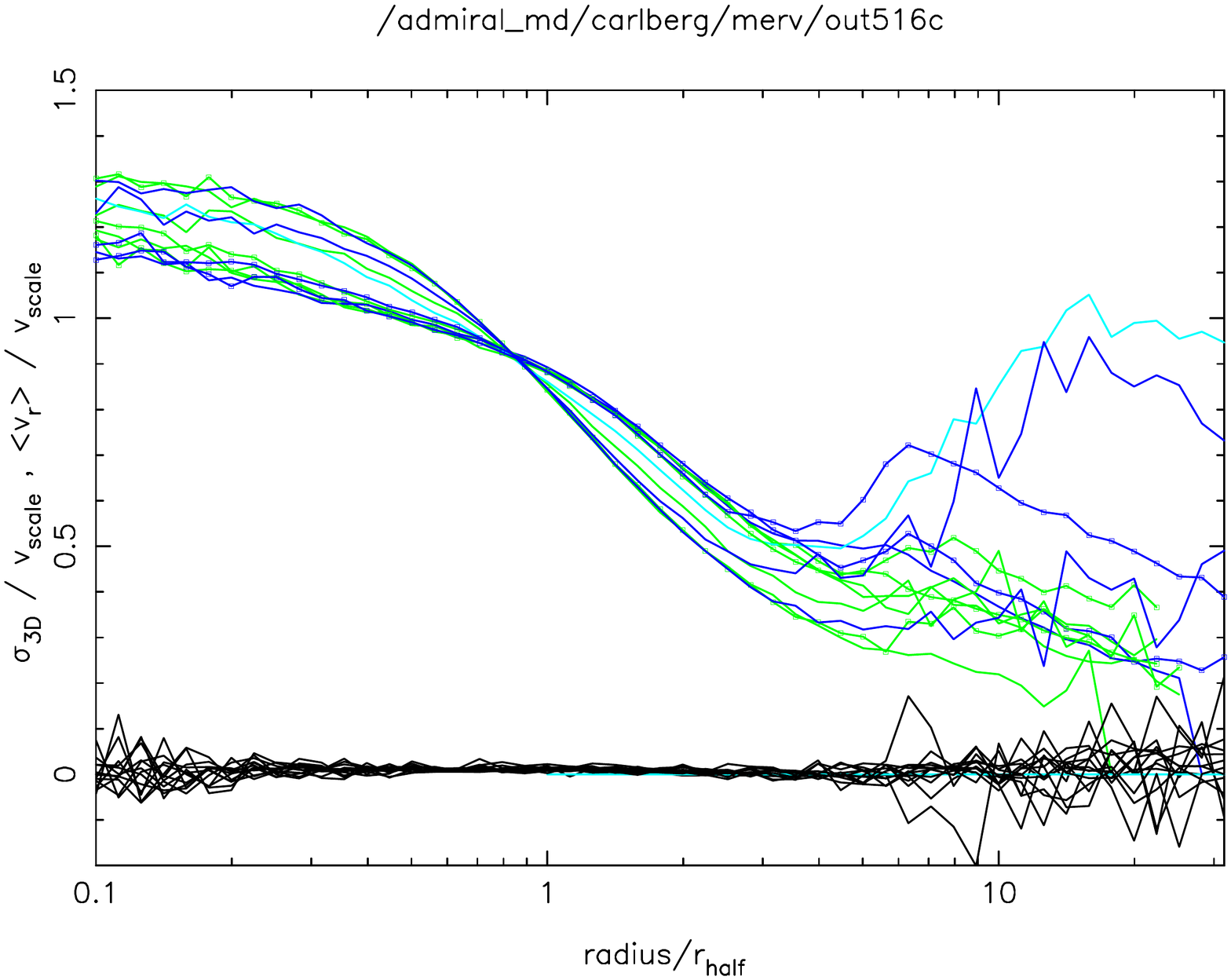}
\end{center}
\caption{The normalized velocity dispersion profile at the final time for the clusters started at sub-halo centers that have remnant dark matter halos with a local dark matter density profile of $r^{-1}$ or steeper within 400 pc. 
}
\label{fig_vcdmax}
\end{figure}

The cluster velocity  dispersion profiles beyond about half the tidal radius have an orbital phase dependence which is a response to the variation in the tidal field around the orbit. The color of the line indicates the radial location of the cluster within the galaxy, with red for the inner clusters and blue for the outer ones. Clusters moving outwards on their orbits have the velocity dispersion curve marked with boxes. Tidal fields will be stronger for clusters closer to the galactic center, and the effect of tides is to give stars an outward velocity increase as the cluster passes the orbital pericenter. The large radii velocity dispersion curves in the top panel of Figure~\ref{fig_vc} are consistent with these ideas, although the range of velocity dispersion is not large.

\begin{figure}
\begin{center}
\includegraphics[angle=0,scale=0.35,trim=30 30 10 70, clip=true]{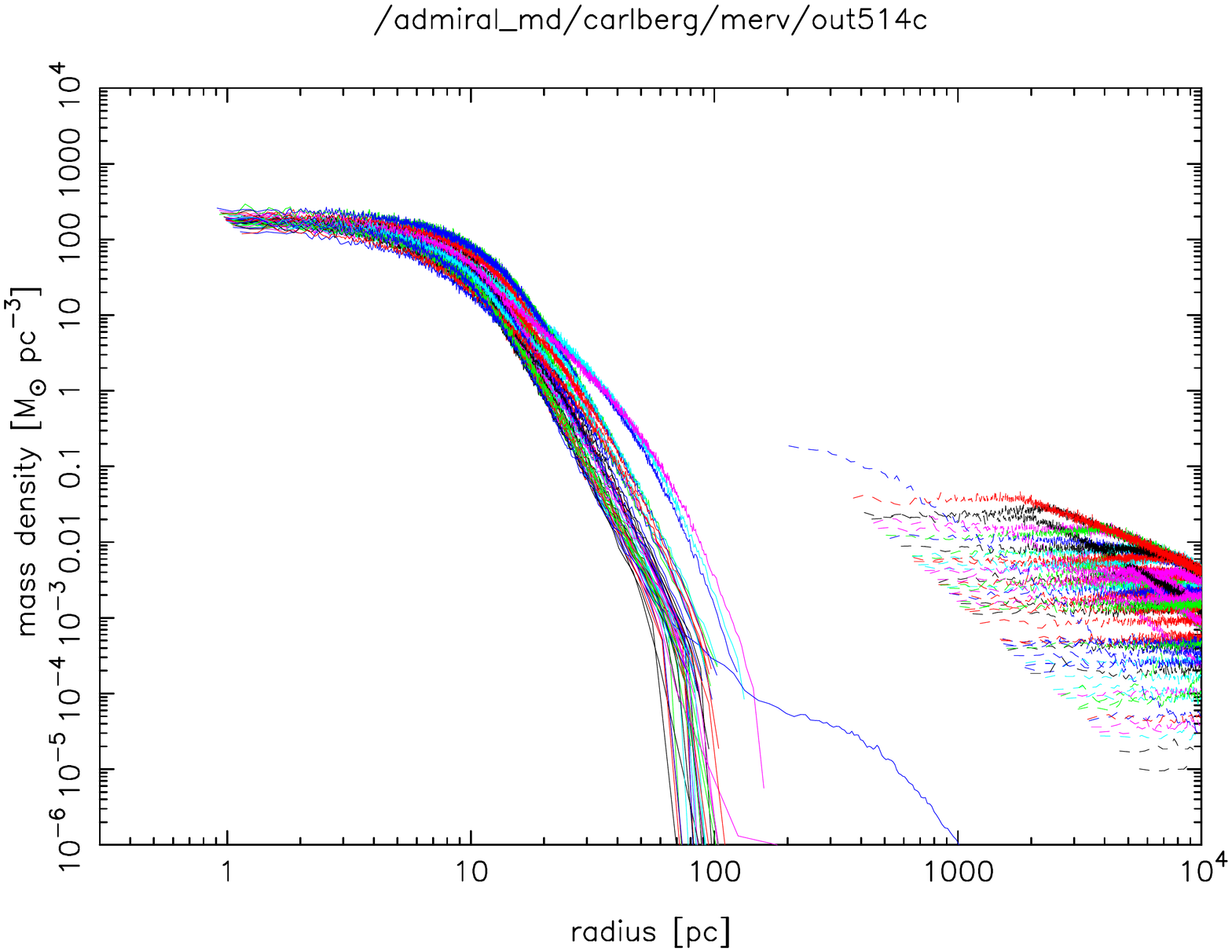}
\includegraphics[angle=0,scale=0.35,trim=30 30 10 70, clip=true]{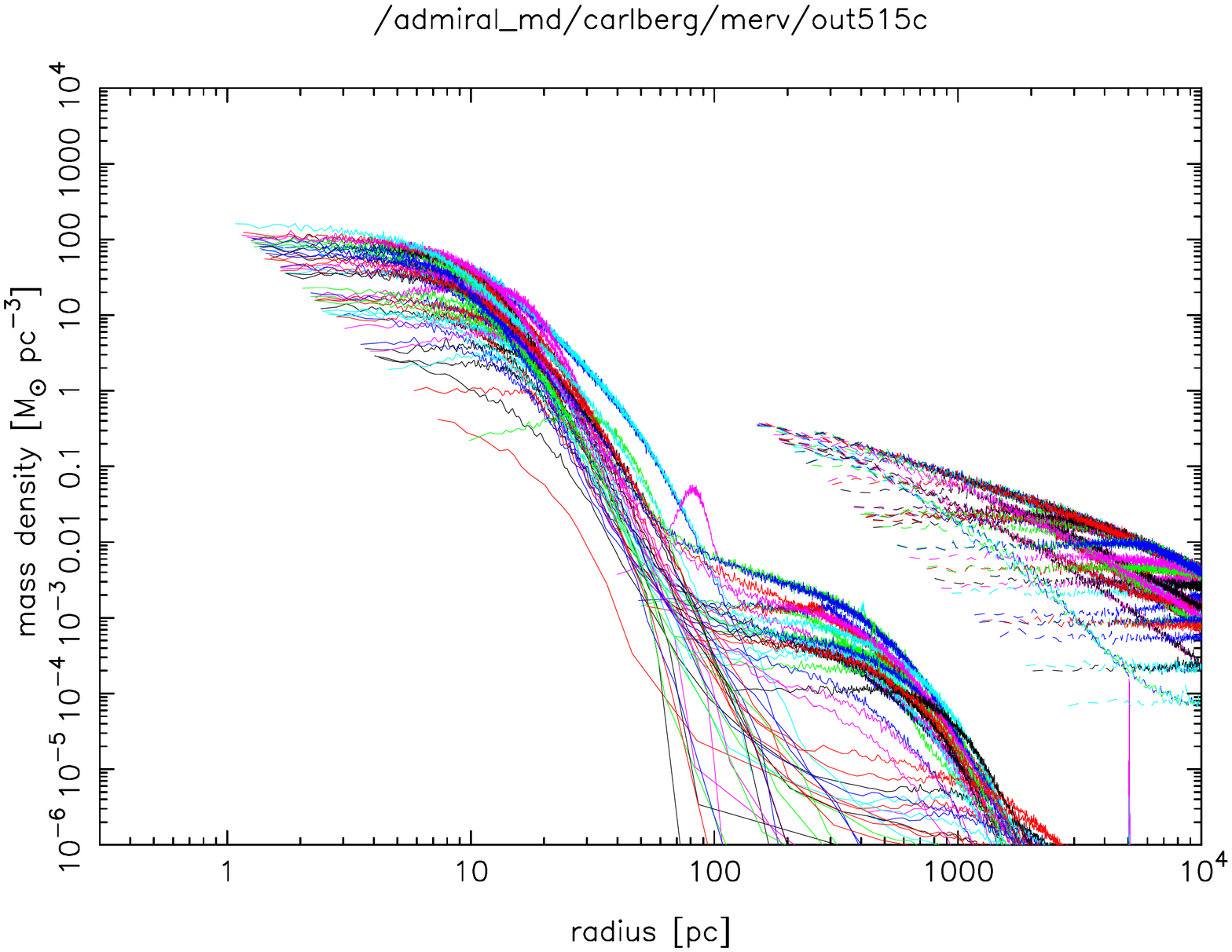}
\includegraphics[angle=0,scale=0.35,trim=30 30 10 70, clip=true]{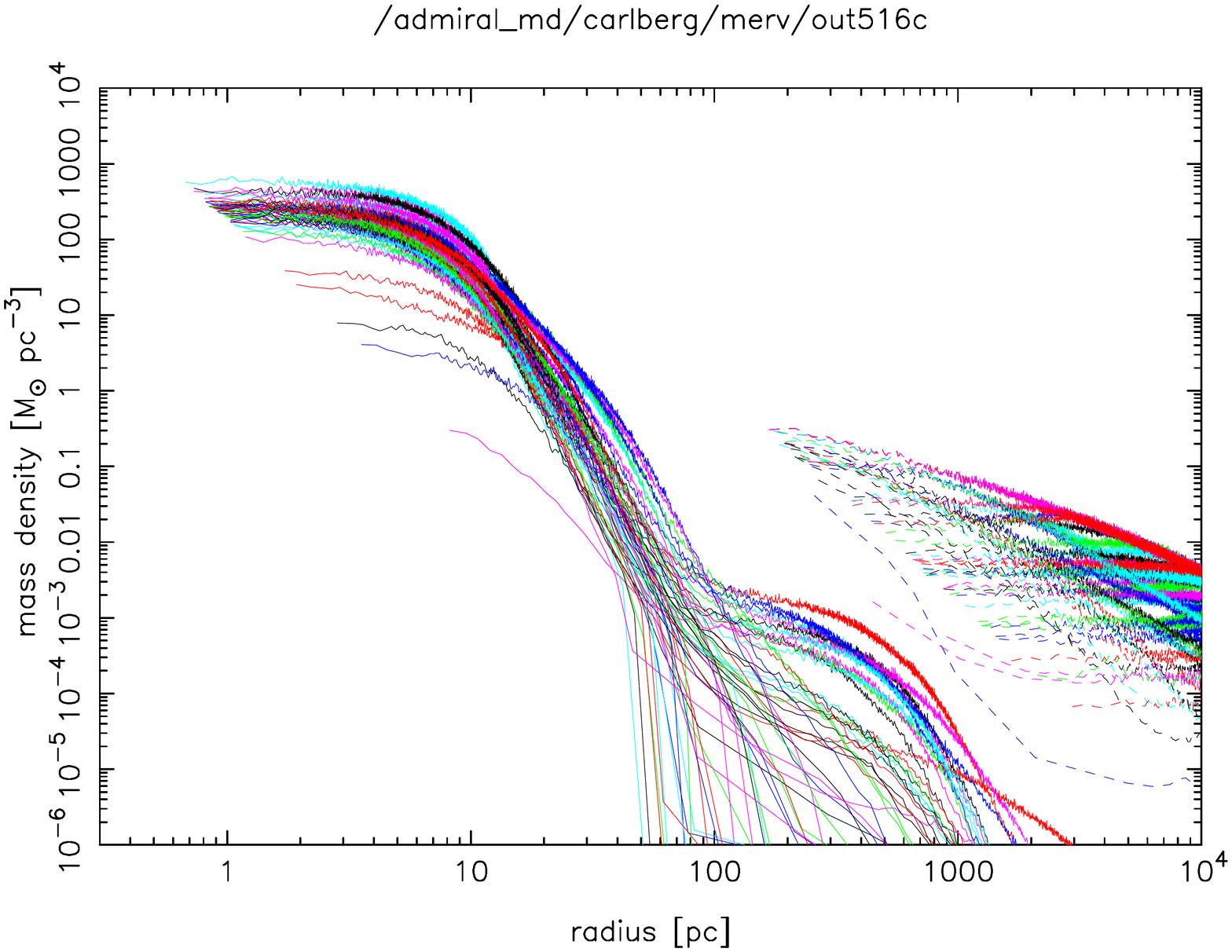}
\end{center}
\caption{The mass density as a function of radius for the same clusters and colors as in Fig.~\ref{fig_vc} and \ref{fig_msd} at the final epoch.  The stellar and dark matter mass profiles are shown as dotted and solid lines, respectively. Both bound and unbound stars are included in the density measurement. Stars beyond the tidal radii of about 100 kpc are being drawn out into a stellar stream. 
}
\label{fig_denend}
\end{figure}

\begin{figure}
\begin{center}
\includegraphics[angle=0,scale=0.35,trim=30 30 10 80, clip=true]{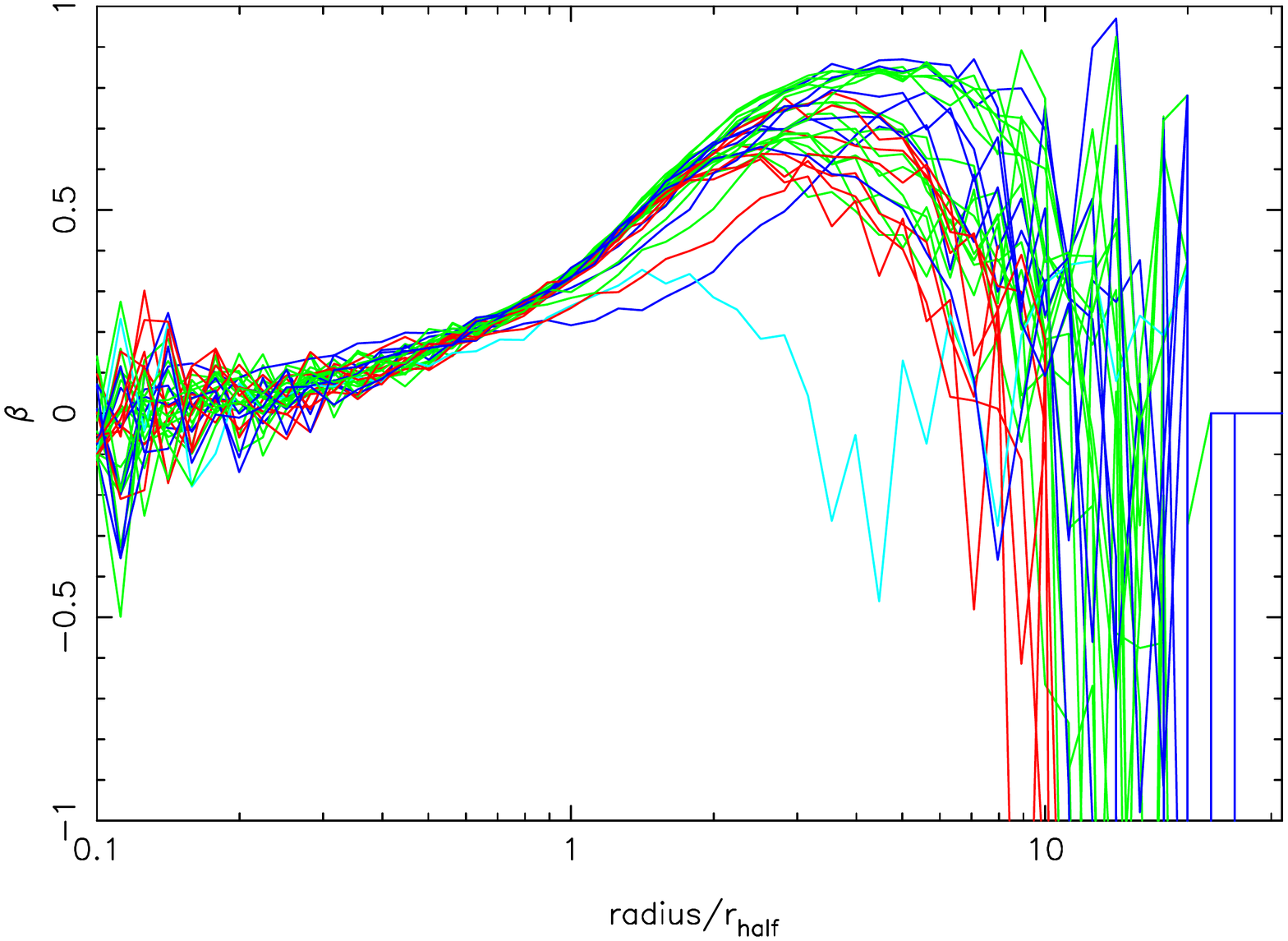}
\includegraphics[angle=0,scale=0.35,trim=30 30 10 80, clip=true]{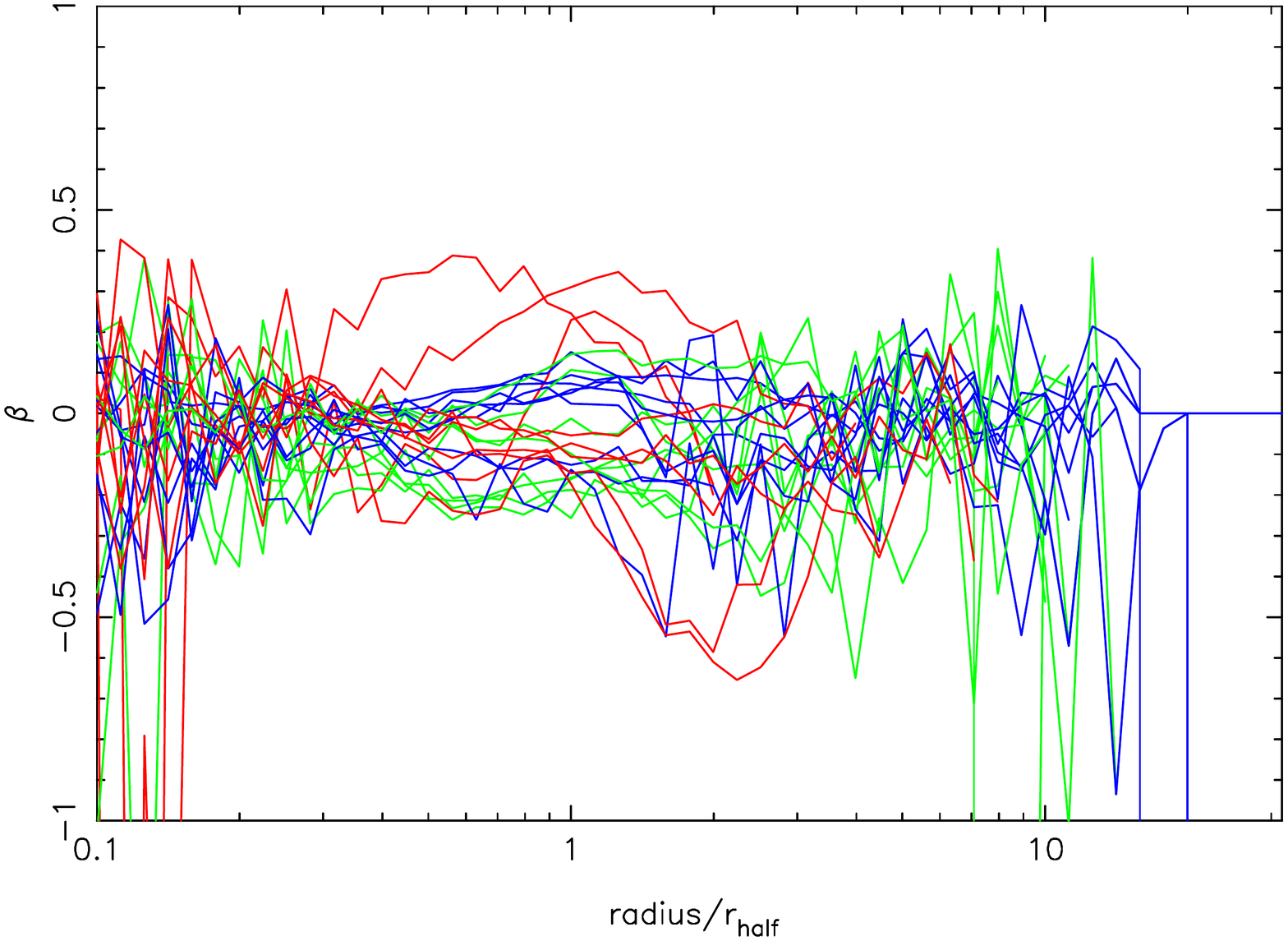}
\includegraphics[angle=0,scale=0.35,trim=30 30 10 80, clip=true]{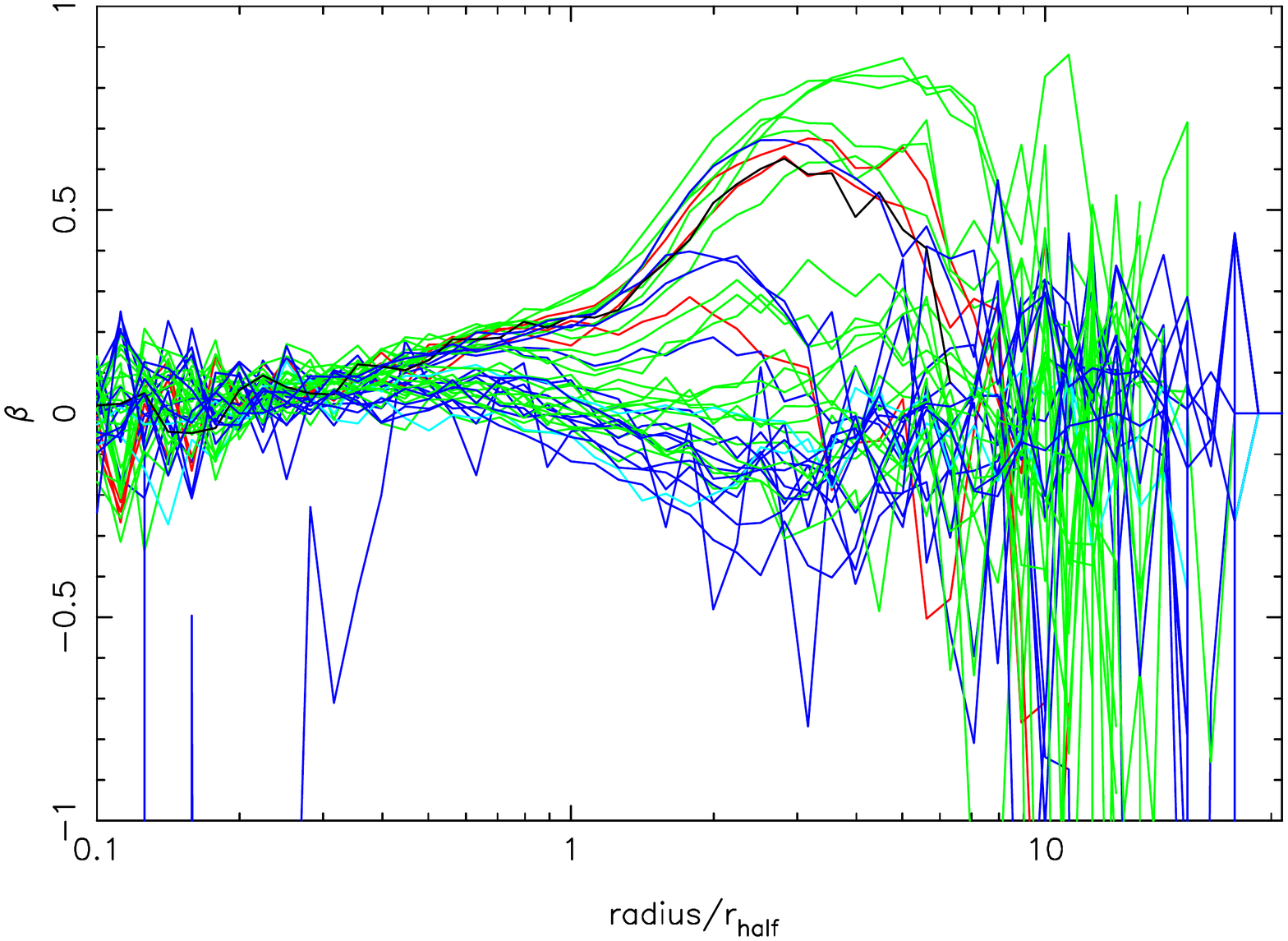}
\end{center}
\caption{The velocity anisotropy parameter, $\beta$, as a function of radius for the same simulation order and colors as in Fig.~\ref{fig_vc} with the same velocity cut applied, plotted within the tidal radius.
}
\label{fig_beta}
\end{figure}

The velocity dispersion profiles of the clusters started at orbital radii of $\approx$0.15 kpc  are shown in the middle panel of Figure~\ref{fig_vc}. For this simulation, all cluster stars are included in the analysis, including the many that are unbound. These clusters show large internal radial velocities, both inward and outward, as the clusters respond to the strong tidal fields around their orbits. These clusters have lost much mass, have large half-mass radii, and are expanding and contracting around their orbits in response to the changing tidal field. Most of the clusters are not in a stationary self-gravitating dynamical equilibrium. On the average these clusters have lost half their mass already, as shown in the middle panel of Figure~\ref{fig_mt} and are continuing to lose mass rapidly. There are few if any clusters like this group in the Milky Way.

The clusters started at the centers of dark matter halos display a range of large radius velocity dispersion profiles, as shown in the bottom panel of Figure~\ref{fig_vc}. The distribution of velocity dispersion at large radius is essentially continuous from being identical to the outer clusters in the top panel,  to an outer rise in velocity dispersion which for a few reaches the velocity dispersion at the half mass radii. The mean radial velocity remains close to zero for all these clusters and they are in a self-gravitating equilibrium. The colors of the lines are related to the distance of the cluster from the center of the halo. The bottom panel shows the velocities have a much smaller rise at large radius if the star cluster is close to the center of the main galactic halo (red and black lines).  The clusters that are most likely to show a rise in velocity dispersion are at galacto-centric radii between 30 and 100 kpc.

The rise in velocity dispersion at large radius can have two sources; either  non-stellar (dark) mass at large radius, or tidally driven velocity variations around the cluster orbit. A measurement of the local dark mass is present is shown in Figure~\ref{fig_msd}. The dark matter particles with velocities less than the escape velocity from the cluster at its cluster-centric radius $r$, $v_{\rm esc}=\sqrt{2GM_c/r}$, are shown in the solid lines, and all dark matter particles are shown in the dashed lines. However, dark matter does not need to be bound to a cluster to have an influence on the outer velocity dispersion profile of the cluster. That is, if there is a  local potential minimum in the dark matter distribution which a local density maximum in the dark matter creates, then high velocity stars at large radius can be bound within the local dark matter sub-halo potential. Figure~\ref{fig_montage} shows the projected dark matter distribution centered on all the star clusters between 3 and 150 kpc of the center of the main galactic halo.  Local density maxima around clusters are identified as those clusters when the ratio of  the dark matter density within 100 pc of the cluster center to the density within the shell between 80 and 400 pc is at least 5, which indicates an average local dark matter density profile of $r^{-1}$ or steeper.  The velocity dispersion profiles of the 13 clusters that meet this criterion are plotted in Figure~\ref{fig_vcdmax}. These clusters have a large radius velocity dispersion profile that is either flat or rising. The  velocity dispersion curves of Figure~\ref{fig_vc} and \ref{fig_vcdmax} are marked with boxes if they are moving outwards on their galactic orbits, with roughly half of the curves with large rises in velocity dispersion being outward moving. Overall, we conclude that rising velocity curves are the result of the cluster being in the remnant of a dark matter sub-halo.

The clusters that are started on $\approx$5 kpc orbits within the sub-halos, the top two panels, do not begin within any bound dark matter, although one or two pick up a few bound particles. The clusters started in the sub-halo centers can retain local remnant dark matter over the evolution of the simulation, leaving a local dark matter mass that is as much as half of the cluster mass or as little as none. Clusters at greater distances from the galactic center (the blue lines) are more likely to retain bound dark matter than inner clusters (green lines). The corresponding mass density distributions of the stars and dark matter particles for the three simulations are displayed in  Figure~\ref{fig_denend}. 

The velocity anisotropy parameter, $\beta=1-\sigma_t^2/(2\sigma_r^2)$, is shown in Figure~\ref{fig_beta}. The clusters started at large sub-halo radii, without local dark matter, develop a velocity anisotropy rise, that is, more radial, as shown in the top panel of Figure~\ref{fig_beta}. The anisotropy peaks in the range of 4-5 half mass radii with values in the range of 0.5 to 0.8. In striking contrast, the star clusters embedded in sub-halo centers (bottom panel) and velocity rises at large radii have little velocity anisotropy at any radius. The clusters with little velocity dispersion rise at large radii behave like clusters with no local dark matter, showing significant radial anisotropy at large radii. The clusters on large orbits with some local dark matter usually have little velocity anisotropy.

\section{Discussion and Conclusions}

Our simulations show that massive, tidally limited, dense star clusters that form in high redshift sub-halos will have surviving members at the present epoch, with half-mass radii and galactic distribution comparable to observed halo clusters. All the simulated clusters lose mass to produce stellar streams, 1-2\% of the mass for star clusters started at sub-halo radii of $\approx$5 kpc, and 10\% for those started at the sub-halo centers. Star clusters started deep in the sub-halos,  $\approx$0.15 kpc, lose  50\% of their mass on the average, with some disrupted completely.  The simulations were initiated with only massive clusters, greater than $4\times 10^5 M_\sun$. Had lower mass clusters been included they would have lost relatively more mass and had a larger fraction of their initial numbers disrupted.  The $\approx$5 kpc, $\approx$0.15 kpc, and centered star clusters are distributed throughout the main halo, with mean radii of 30, 5, and 16 kpc in the main halo. 

The current epoch internal kinematics of the star clusters beyond 3 half mass radii are strikingly dependent on the high redshift formation location. The clusters started at large sub-halo radii have a velocity dispersion that declines to 15-25\% of the central value at 10-20 half mass radii, but never reaches zero. The star clusters started at sub-halo centers often remain in the center of a remnant dark matter sub-halo. These clusters end with outer velocity dispersion profiles that go from declining to an outer rise in velocity dispersion up to the value at the half-mass radius.   Clusters with remnant local dark matter sub-halos show a rise in their velocity dispersion profile at large cluster-centric radius.

These simulations are useful templates for comparison to the growing amount and accuracy of stellar velocities of globular clusters stars at large radii to test for the presence of local dark halos.

\begin{acknowledgements}
An anonymous referee provided helpful comments. 
This research was supported by NSERC of Canada. Computations were performed on the niagara supercomputer at the SciNet HPC Consortium. SciNet is funded by: the Canada Foundation for Innovation; the Government of Ontario; Ontario Research Fund - Research Excellence; and the University of Toronto.
\end{acknowledgements}

\software{Gadget4: \citet{Gadget4}, Amiga Halo Finder: \citep{2004MNRAS.351..399G,2009ApJS..182..608K}}

\bibliography{GCdark}{}
\bibliographystyle{aasjournal}

%\listofchanges
\end{document}